%% file: noiseAssumption.tex
\documentclass[a4paper]{quantumarticle}
\pdfoutput=1
\usepackage{float, verbatim, amsmath, amssymb, ulem, color}
\usepackage[numbers,sort&compress]{natbib}

\title{The Discrete Noise Approximation in Quantum Circuits}
\author{Keith R.\ Fratus}
\author{Juha Leppäkangas}
\author{Michael Marthaler}
\author{Jan-Michael Reiner}
\affiliation{HQS Quantum Simulations GmbH, Rintheimer Straße 23, 76131 Karlsruhe, Germany}

\begin{document}

\maketitle

\input{Sections/abstract}

\input{Sections/intro}

\input{Sections/lind}

\input{Sections/resSum}

\input{Sections/justifGiven}

\input{Sections/sepNoiseAcc}

\input{Sections/sepNoiseExamp}

\input{Sections/applications}

\input{Sections/conclusion}

\input{Sections/acknowledgment}

\bibliographystyle{unsrtnat}
\bibliography{Sections/refs}

\newpage

\appendix

\input{Sections/appSum}

\input{Sections/appCommuteGate}

\input{Sections/appSepDeriv}

\input{Sections/appPSD}

\input{Sections/appNumerics}

\input{Sections/appSteady}

\input{Sections/appGateExamp}

\end{document}

%% file: Sections/abstract.tex
\begin{abstract}

When modeling the effects of noise on quantum circuits, one often makes the assumption that these effects can be accounted for by individual decoherence events following an otherwise noise-free gate. In this work, we address the validity of this model. We find that under a fairly broad set of assumptions, this model of individual decoherence events provides a good approximation to the true noise processes occurring on a quantum device during the implementation of a quantum circuit. However, for gates which correspond to sufficiently large rotations of the qubit register, we find that the qualitative nature of these noise terms can vary significantly from the nature of the noise at the underlying hardware level. The bulk of our analysis is directed towards analyzing what we refer to as the separability ansatz, which is an ansatz concerning the manner in which individual quantum operations acting on a quantum system can be approximated. In addition to the primary motivation of this work, we identify several other areas of open research which may benefit from the results we derive here. 

\end{abstract}

%% file: Sections/intro.tex
\section{Introduction}
\label{sec:introduction}

While progress is being made in the field of quantum hardware, the days of fully error-corrected quantum computation still lie over the horizon. For this reason, considerable effort has been put into understanding what can be accomplished with Noisy Intermediate Scale Quantum (NISQ) devices, and to what extent noise affects their performance \cite{Barthi2022}. When describing a quantum computation at the circuit level, the effects of noise are often modeled as an approximation at the gate level,
\begin{equation}
\mathcal{G} \approx \mathcal{N}_{G} G,
\end{equation}
where $\mathcal{G}$ is a quantum operation which results from the application of an imperfect gate, $G$ represents the desired unitary gate operation which would ideally be implemented in the absence of noise, and $\mathcal{N}_{G}$ is a discrete decoherence event. In this work, we would like to address the accuracy of this model.

We show that given a reasonable set of assumptions regarding how quantum gates are implemented on a quantum device, along with a similarly reasonable set of assumptions concerning how this device is affected by noise, to lowest order in the noise strength we can always write
\begin{equation}
\mathcal{N}_{G} \to \exp \left ( t_{G} \mathcal{L}^{G}_{N} \right ),
\end{equation}
where $t_{G}$ is the time required to implement the gate, and $\mathcal{L}^{G}_{N}$ is a term of pure Lindblad form, with no coherent part. We emphasize, however, that the term $\mathcal{L}^{G}_{N}$ may deviate significantly from the original noise acting on the hardware, depending on the precise nature in which the gate is implemented. In particular, we find that strong deviations from the original hardware noise do not require the effects of noise to be particularly ``large'' with respect to the coherent time evolution of the gate, and that likewise it is not necessarily possible to ameliorate such deviations by applying the action of the gate sufficiently ``quickly'' with respect to the time scale of the underlying environmental noise. The only requirement for such deviation to occur is that the rotations being applied to the qubit register during the implementation of the gate are sufficiently large.

In order to argue for such a claim, the bulk of our work in this paper is focused on analyzing what we refer to as the \textit{separability ansatz}, which is an ansatz concerning the manner in which individual quantum operations on a hardware device can be approximated. In particular, we consider a quantum system with density matrix $\rho$ being evolved according to a Lindblad-like time evolution generator,
\begin{equation}
\mathcal{L} = \mathcal{L}_{H} + \mathcal{L}_{D},
\end{equation}
where $\mathcal{L}_{H}$ is some coherent Hamiltonian evolution, and $\mathcal{L}_{D}$ describes the effect of noise on the system during this Hamiltonian evolution. If one wishes to evaluate the quantum operation $\Phi$ corresponding to an evolution of the system under $\mathcal{L}$ for a time $t_{\text{op}}$,
\begin{equation}
\frac{d}{dt} \rho = \mathcal{L} \left [ \rho \right ] ~ \to ~ \rho \left ( t_{\text{op}} \right ) = \Phi \left [ \rho_{0} \right ],
\end{equation}
then to lowest order in the noise strength, we find that it is always possible to write
\begin{equation}
\Phi \approx \exp \left ( t_{\text{op}} \mathcal{L}_{S} \right ) \phi,
\end{equation}
where  $\mathcal{L}_{S}$ is an effective, \textit{separated} noise term, which is purely Lindblad-like, with no Hamiltonian part, and $\phi$ is the quantum map corresponding to the purely Hamiltonian evolution, in the absence of any noise,
\begin{equation}
\frac{d}{dt} \rho = \mathcal{L}_{H} \left [ \rho \right ] ~ \to ~ \rho \left ( t_{\text{op}} \right ) = \phi \left [ \rho_{0} \right ].
\end{equation}
While the form of the effective noise term $\mathcal{L}_{S}$ may differ significantly from that of the original noise term $\mathcal{L}_{D}$, the overall noise strength of $\mathcal{L}_{S}$ remains the same as $\mathcal{L}_{D}$. Not only do our results concerning this ansatz provide justification for the model of gate noise considered here in the context of quantum computation, but they may also hint at the possibility of developing novel methods for studying the subject of Lindblad time evolution in general.

The outline of this paper is as follows: In Section~\ref{sec:lind} we review the basic properties of the Lindblad equation and introduce some notation which will be used to state our results. In Section~\ref{sec:resSum} we state the main results of our analysis concerning the separability ansatz, while in Section~\ref{sec:justifGiven} we explain how our assumptions regarding hardware noise, together with the separability ansatz, lead to our model of gate noise. In Section~\ref{sec:sepNoiseAcc} we numerically evaluate the accuracy of the separated noise approximation, and in Section~\ref{sec:sepNoiseExamp} we give several concrete examples of separated noise terms, and discuss some generic features of separated noise terms. In Section~\ref{sec:applications} we discuss several potential applications of our work, and then we conclude in Section~\ref{sec:conclusion}. In Appendix~\ref{sec:appCommuteGate} we derive various results involving the commutators of coherent and incoherent quantum operations, which we use in the derivation of our main results concerning the separability ansatz in Appendix~\ref{sec:appSepDeriv}. In Appendix~\ref{sec:appPhys} we discuss the necessary conditions for the separated noise term to be physical, while in Appendix~\ref{sec:appNumerics} we elaborate in more detail on how the separated noise can be computed in practice in the time-independent case, using both numerical and analytic methods. In Appendix~\ref{sec:appSteady} we derive some results concerning certain steady state behavior of the separated noise, while in Appendix~\ref{sec:appGateExamp} we elaborate further on some specific examples of gate noise terms.

%% file: Sections/lind.tex
\section{The Lindblad Equation and its Associated Quantum Maps}
\label{sec:lind}

The most general linear, Markovian evolution on the space of density matrices can be written \cite{RevModPhys.88.021002}
\begin{equation}
\begin{split}
& \frac{d}{dt}\rho \\ = ~ & \mathcal{L} \left [ \rho  \right ] \\ = ~ & \mathcal{L}_{H}\left [ \rho  \right ] + \mathcal{L}_{D}\left [ \rho  \right ] \\
= ~ & -i \left [ H, \rho \right ] + \sum_{n, m} \Gamma^{D}_{nm} \left ( A_{n} \rho A_{m}^{\dagger} - \frac{1}{2} \left \{ A_{m}^{\dagger} A_{n}, \rho \right \}  \right ).
 \end{split}
\end{equation}
The first term in this equation corresponds to unitary evolution under a Hamiltonian $H$, while the second term, described by the Hermitian \textit{rate matrix} $\Gamma$, allows for decoherence of the density matrix, and is often associated with external noise acting on the system of interest. The operators $ \left \{ A_{n} \right \} $ represent a basis for the space of all traceless operators on the relevant Hilbert space $\mathcal{H}$ of dimension $\mathcal{D}$.

When the Hamiltonian and rate matrix are time-independent, and the rate matrix is positive semi-definite (PSD), we recover the Lindblad equation. The Lindblad equation can be shown to be the most general time-homogeneous, Markovian map which preserves the statistical interpretation of the density matrix \cite{lindblad}. When this assumption of time-independence is relaxed, the necessary conditions for preserving the statistical interpretation of the density matrix are still not fully understood \cite{RevModPhys.88.021002}. A sufficient (but not always necessary) condition is that the rate matrix remains PSD at all times. In our work, we will often assume that the rate matrix is PSD at all times, although our results derived here do not specifically require this. More discussion of the Lindblad equation can be found in \cite{lidar, breuer2002theory}.

Since the object $\mathcal{L}$ acts linearly on an operator to return another operator, it is naturally interpreted as a super-operator acting on the vector space of (regular) operators, whose matrix elements with respect to the basis $\left \{ A_{n} \right \}$ are defined as
\begin{equation}
\mathcal{L} \left [  A_{n} \right ] = \sum_{m} \mathcal{L}_{mn}  A_{m}.
\end{equation}
The object $\mathcal{L}$ naturally acts on the density matrix $\rho$, which must necessarily have a component along the identity operator. However, it is straightforward to verify that the action of $\mathcal{L}$ is in fact closed on the space of traceless operators $ \left \{ A_{n} \right \} $, so it makes sense to talk about matrix elements of $\mathcal{L}$ in this restricted space. One should be careful, however, not to confuse the matrix elements $\mathcal{L}_{mn}$ with the elements $\Gamma^{D}_{nm}$ of the rate matrix, as their interpretation is fundamentally different. If we have defined an inner product on this space of operators, we can use it to compute the matrix elements of $\mathcal{L}$ according to
\begin{equation}
\mathcal{L}_{mn} = \sum_{p} \eta^{-1}_{mp} \langle \langle A_{p} || \mathcal{L} || A_{n} \rangle \rangle = \sum_{p} \eta^{-1}_{mp}\langle \langle A_{p} || \mathcal{L} \left [  A_{n} \right ] \rangle \rangle,
\end{equation}
where the matrix of inner products
\begin{equation}
\eta_{mp} =  \langle \langle A_{m} || A_{p} \rangle \rangle
\end{equation}
is invertible by virtue of the basis $\left \{ A_{n} \right \}$ being complete, and where we have used the double bracket notation to distinguish the inner product on operators from the usual inner product on vectors. We can also use such an inner product to define the adjoint of this (or any other) super-operator in the usual way,
\begin{equation}
\langle \langle A || \mathcal{L}^{\ddagger} || B \rangle \rangle = \langle \langle  \mathcal{L} \left [ A \right ] || B \rangle \rangle,
\end{equation}
where we use the double dagger notation to avoid confusion with the adjoint of a (regular) operator. For our purposes, we will choose to define the usual Frobenius inner product
\begin{equation}
\langle \langle A || B \rangle \rangle ~ \equiv ~ \frac{1}{\mathcal{D}}\text{Tr} \left [ A^{\dagger}B \right ],
\end{equation}
and will always choose our basis $\left \{ A_{n} \right \}$ so that it is orthonormal with respect to this inner product,
\begin{equation}
\langle \langle A_{m} || A_{n} \rangle \rangle ~ = ~ \frac{1}{\mathcal{D}}\text{Tr} \left [ A^{\dagger}_{m}A_{n}\right ]  ~=~ \delta_{mn},
\end{equation}
with the matrix $\eta$ thus reducing to the identity.

Common choices for the basis $ \left \{ A_{n} \right \} $ include the (properly normalized) generalized Gell-Mann matrices \cite{Bertlmann_2008}, or, since we are typically interested in collections of qubits, the set of all possible products of Pauli operators on $k$ qubits,
\begin{equation}
A_{\alpha} \equiv \bigotimes_{i=1}^{k} \sigma_i^{\alpha_i}.
\end{equation}
If we have some new basis of traceless operators $\left \{ B_{n} \right \}$ defined according to
\begin{equation}
B_{n} = \sum_{p} V_{pn} A_{p}
\end{equation}
then one can verify by direct insertion into the Lindblad equation that the rate matrix in this new basis is related to the old rate matrix by
\begin{equation}
\Gamma^{D;A} = V \Gamma^{D;B} V^{\dagger}.
\end{equation}
where we explicitly emphasize the basis dependence of the rate matrix. In the case that this new basis is defined in terms of the old one by some super-operator $\pi$,
\begin{equation}
B_{n} = \pi \left [ A_{n} \right ],
\end{equation}
then tracing both sides of the defining equation for $V$ against the operator $A_{m}^{\dagger}$ reveals that it is simply equal to the matrix of the super-operator $\pi$ evaluated in the original basis,
\begin{equation}
V_{mn} = \langle \langle A_{m} || \pi || A_{n} \rangle \rangle.
\end{equation}
We note that in order for the new basis to remain orthonormal, the matrix $V$ must be a unitary matrix, corresponding to a super-operator $\pi$ which is a unitary transformation with respect to the Frobenius inner product. In this case, we have
\begin{equation}
V^{\dagger} = V^{-1} ~\Rightarrow~ \Gamma^{D;A} = V \Gamma^{D;B} V^{-1}.
\end{equation}
which is to say that the rate matrix transforms like a tensor under such a change of basis (there is a similar sense in which the Hamiltonian transforms like a vector, but we will not need this fact for our analysis).

A particularly relevant choice of basis for the incoherent part of the Lindblad equation is given by
\begin{equation}
L_{i} = \sum v_{i}^{(n)} A_{n}.
\end{equation}
where the vectors $\left \{ \vec{v}_{i} \right \}$ are the (normalized) eigenvectors of the rate matrix (as expressed in the basis $ \left \{ A_{n} \right \} $). Because the rate matrix is Hermitian, these eigenvectors will always constitute a complete basis, with real eigenvalues $\{\gamma_{i}\}$, and in the case that it is also positive semi-definite, these eigenvalues will be non-negative. This choice of basis leads to the more common diagonal form,
\begin{equation}
\mathcal{L}_{D}\left [ \rho  \right ] = \sum_{i} \gamma^{D}_{i} \left [ L_{i} \left ( \rho \right )L_{i}^{\dagger} - \frac{1}{2} \left \{ L_{i}^{\dagger} L_{i}, \rho \right \} \right ],
\end{equation}
The eigenvalues $\{\gamma^{D}_{i}\}$ correspond to the physical decay rates of the system under the effects of decoherence.  Throughout this work, we will often refer to the ``total noise strength'' of a Lindblad term, which for our purposes we define as
\begin{equation}
\overline{\gamma} \equiv \text{Tr} \left [ \Gamma \right ].
\end{equation}
Note that this object is equal to the sum of the eigenvalues of the rate matrix, and does not depend on the choice of basis $\left \{ A_{n} \right \}$, so long as it remains orthonormal. This object is useful for comparing the cumulative effects of noise in a basis-independent manner between two Lindblad terms which differ in their structure.

The solution to the Lindblad equation is formally given
\begin{equation}
\rho \left ( t_{\text{op}} \right ) = \Phi \left [ \rho_{0} \right ],
\end{equation}
with the quantum operation $\Phi$ given according to
\begin{equation}
\Phi = \mathcal{T} \left \{ \exp \left ( \int_{0}^{t_{\text{op}}} \mathcal{L} \left ( t \right ) dt \right ) \right \}
\end{equation}
where $\mathcal T$ is the time-ordering symbol. We will often be interested in the dynamics under the purely coherent portion of the dynamics, which we define
\begin{equation}
\phi = \mathcal{T} \left \{ \exp \left ( \int_{0}^{t_{\text{op}}} \mathcal{L}_{H} \left ( t \right ) dt \right ) \right \}.
\end{equation}
In the purely coherent case, this quantum map can alternatively be written in terms of a unitary matrix $U$,
\begin{equation}
\label{eq:adrel}
\phi \left [ \rho \right ] = U \rho U^{\dagger},
\end{equation}
with the usual definition,
\begin{equation}
U = \mathcal{T} \left \{ \exp \left ( -i\int_{0}^{t_{\text{op}}} H \left ( t \right ) dt \right ) \right \}.
\end{equation}
It is straightforward to verify that with our particular choice of inner product, the map $\phi$ is itself a unitary super-operator,
\begin{equation}
\phi^{\ddagger} = \phi^{-1}.
\end{equation}
We emphasize, however, that not all unitary super-operators can be written in the form of equation (\ref{eq:adrel}) (which can be seen by noting that the number of unitary super-operators is necessarily larger than the number of unitary operators). 

Many of the results of our analysis will involve matrix elements of the super-operator $\phi$,
\begin{equation}
M_{mn} = \langle \langle A_{m} || \phi || A_{n} \rangle \rangle = \frac{1}{\mathcal{D}}\text{Tr} \left [ A_{m}^{\dagger} U  A_{n} U^{\dagger} \right ].
\end{equation}
A useful form of this matrix element is found by introducing the notation of the adjoint action under an operator $A_{p}$ as 
\begin{equation}
\text{ad}_{A_{p}} \equiv \left [ A_{p}, \cdot \right ].
\end{equation}
The matrix elements of such an action are given according to
\begin{equation}
\begin{split}
\left ( \Omega_{p} \right)_{mn} & = \langle \langle A_{m} || \text{ad}_{A_{p}} || A_{n} \rangle \rangle \\ & = \frac{1}{\mathcal{D}}\text{Tr} \left [ A_{m}^{\dagger} \left [ A_{p}, A_{n} \right ] \right ] \equiv -i g_{pmn}.
\end{split}
\end{equation}
If we expand the Hamiltonian according to
\begin{equation}
H = \sum_{p} H_{p} A_{p}
\end{equation}
then we can further define the matrix element
\begin{equation}
\Omega_{mn} = \langle \langle A_{m} || \text{ad}_{H} || A_{n} \rangle \rangle = -i \sum_{p} H_{p} g_{pmn}.
\end{equation}
With this notation, we can write the quantum map $\phi$ according to
\begin{equation}
\phi = \mathcal{T} \left \{ \exp \left ( -i\int_{0}^{t_{\text{op}}} \text{ad}_{H} \left ( t \right ) dt \right ) \right \},
\end{equation}
Since $M$ is the matrix representation of $\phi$, while $\Omega$ is the matrix representation of $\text{ad}_{H}$, we can thus write
\begin{equation}
M = \mathcal{T} \left \{ \exp \left ( -i\int_{0}^{t_{\text{op}}} \Omega \left ( t \right ) dt \right ) \right \}.
\end{equation}
The fact that this relationship is not complicated by the presence of the time-ordering symbol is argued in Appendix~\ref{sec:appCommuteGate}. We note that the matrix $\Omega$ is Hermitian, while the matrix $M$ is unitary.

In the case that the $\left \{ A_{n} \right \}$ are all chosen to be Hermitian, it is natural to interpret them as forming a basis of the Lie Algebra $\mathfrak{su}(\mathcal{D})$. In this case, the $g_{pmn}$ are nothing other than the structure constants $f_{pmn}$ of the algebra, and since $\mathfrak{su}(\mathcal{D})$ is a semi-simple algebra, they can be taken to be real and totally anti-symmetric, with no need to distinguish upper from lower indices \cite{wybourne1974classical}. Since the structure constants of $\mathfrak{su}(\mathcal{D})$ are known in closed form \cite{bossion2021general}, such a choice can help with the calculation of the matrix $\Omega$. However, we emphasize that none of the results of our analysis require such a choice, and it is important to remember that in the general case, the $g_{pmn}$ are not necessarily real or totally anti-symmetric.

%% file: Sections/resSum.tex
\section{Summary of the Main Results Concerning the Separability Ansatz}
\label{sec:resSum}

We state in this section the main results of our analysis concerning the separability ansatz. In the process of analyzing this ansatz, we must study the manner in which a Lindblad noise term is modified after being commuted past a coherent quantum operation. We quote here these results as a corollary to our main result. The proof of these results can be found in Appendix~\ref{sec:appCommuteGate}. In particular, we wish to solve the equation
\begin{equation}
\phi e^{\mathcal{L}_{D}} = e^{\mathcal{L}_{L}} \phi 
\end{equation}
where $\mathcal{L}_{D}$ is the original Lindblad dissipator, $\mathcal{L}_{L}$ is the transformed dissipator, and $\phi$ is the coherent quantum operation, generated by some coherent time-evolution operator $\mathcal{L}_{H}$. A formal solution to this equation is given by
\begin{equation}
\mathcal{L}_{L} = \phi \mathcal{L}_{D} \phi^{-1}.
\end{equation}
This transformed object is in fact still a pure Lindblad dissipator. It can be described by retaining the original rate matrix $\Gamma^{D}$ of $\mathcal{L}_{D}$ and making the transformation
\begin{equation}
A_{n} \to B_{n} = \phi \left [ A_{n} \right ] =  U A_{n} U^{\dagger}
\end{equation}
to our basis $\{A_{n}\}$. That is to say,
\begin{equation}
\Gamma^{L;B} = \Gamma^{D;A}
\end{equation}
If we choose to work strictly in the original basis, the rate matrix of the new dissipator can be described in terms of a transformation acting on the original rate matrix, given according to
\begin{equation}
\label{eq:Mrot}
\Gamma^{L} = M\Gamma^{D}M^{\dagger},
\end{equation}
where $M$ is the matrix of the quantum map $\phi$. Since the matrix $M$ is unitary, the transformed rate matrix will have precisely the same spectrum as the original rate matrix, and thus it will be PSD and Hermitian whenever the original rate matrix has these properties (and will also have the same total noise strength). In other words, the physical nature of the noise is preserved under this transformation. If we had instead considered the result of commuting a Lindblad term to the right past a coherent quantum operation, we would find a similar transformation, given according to
\begin{equation}
\Gamma^{R} = M^{\dagger}\Gamma^{D}M.
\end{equation}

We now proceed to the main results of our analysis concerning the separability ansatz, which is the claim that to lowest order in the noise strength, it is always possible to write
\begin{equation}
\Phi \approx \exp \left ( t_{\text{op}}\mathcal{L}_{S} \right ) \phi
\end{equation}
where  $\mathcal{L}_{S}$ is an effective, ``separated'' noise term, which is purely Lindblad, with no Hamiltonian part. These results are derived in Appendix~\ref{sec:appSepDeriv}. The separated noise is found by performing a linear transformation on the underlying hardware noise,
\begin{equation}
\mathcal{L}_{S} = \frac{1}{ t_{\text{op}}} \phi \left [ \int_{0}^{t_{\text{op}}} \phi^{-1} \left ( s \right ) \mathcal{L}_{D} \left ( s \right )  \phi \left ( s \right ) ds \right ] \phi^{-1},
\end{equation}
where $\phi \left ( s \right )$ (with an explicit dependence on the parameter $s$) represents the coherent evolution evaluated up to some intermediate time,
\begin{equation}
\phi\left ( s \right )  = \mathcal{T} \left \{ \exp \left ( \int_{0}^{s} \mathcal{L}_{H} \left ( t \right ) dt \right ) \right \}.
\end{equation}
The rate matrix of this separated noise is similarly given as a linear transformation on the rate matrix of the underlying hardware noise,
\begin{equation}
\label{eqn:mainGamma}
\Gamma^{S} = \frac{1}{ t_{\text{op}}} M  \left [ \int_{0}^{t_{\text{op}}} M^{\dagger} \left ( s \right ) \Gamma^{D} \left ( s \right ) M \left ( s \right ) ds \right ] M^{\dagger},
\end{equation}
where $M \left ( s \right )$ (with an explicit dependence on the parameter $s$) is defined similarly to before, with the time integration in the exponent evaluated up to some intermediate time,
\begin{equation}
\label{eqn:mainM}
M \left ( s \right ) \equiv \mathcal{T} \left \{ \exp \left ( -i \int_{0}^{s} \Omega \left ( t \right ) dt \right )   \right \}.
\end{equation}

In order to evaluate this expression as a practical matter, we define the quantity
\begin{equation}
Q \left ( s \right ) \equiv s \Gamma^{S} \left ( s \right ),
\end{equation}
where the parameter $s$ represents some intermediate time less than $t_{\text{op}}$. We also define
\begin{equation}
\xi \equiv -i \text{ad}_{\Omega} = -i  \left [ \Omega, \cdot \right ]
\end{equation}
to indicate the infinitesimal unitary transformation generated under the adjoint action of $\Omega.$ With this notation in place, we can state a differential equation which defines the separated noise,
\begin{equation}
\label{eqn:mainDiff}
\frac{dQ}{ds} =  \xi \left [ Q \right ] + \Gamma^{D} ~;~ Q\left ( s = 0 \right ) = 0. 
\end{equation}
This differential equation can be integrated from zero up through $t_{\text{op}}$ in order to find $Q \left ( t_{\text{op}} \right )$, which then allows us to find the rate matrix of the separated noise.

In the time-independent case, we note that
\begin{equation}
M \left ( s \right ) = \exp \left ( -i \Omega s \right ).
\end{equation}
Using this fact and performing some minor rearrangement of our integral expression, the transformation of the separated noise simplifies to
\begin{equation}
\Gamma^{S} = \frac{1}{ t_{\text{op}}}  \int_{0}^{t_{\text{op}}} e^{ -i \Omega s}\Gamma^{D} e^{ +i \Omega s} ds.
\end{equation}
In this time-independent case, we can alternatively write the separated noise in terms of a function of the transformation $\xi$, defined according to the series expansion
\begin{equation}
\Gamma^{S} =  \mathcal{K}  \left [ \Gamma^{D} \right ] ~;~ \mathcal{K} = \sum_{m=0}^{\infty} \frac{t_{\text{op}}^{m}}{(m+1)!} \xi^{m}.
\end{equation}
This power series can be summed exactly, such that
\begin{equation}
\mathcal{K} ~=~ \frac{e^{t_{\text{op}} \xi}-\mathcal{I}}{t_{\text{op}} \xi},
\end{equation}
where $\mathcal{I}$ is the identity acting on the space of rate matrices. This expression can be useful for finding the separated noise whenever we can diagonalize the transformation $\xi$, which we will discuss in more detail when giving concrete examples of separated noise. 

We emphasize that even though $\Gamma^{S}$ depends on the parameter $t_{\text{op}}$, the equation of motion defining the separated noise dynamics is still time-independent. That is to say, the evolution of the system under the separated noise dynamics is defined by
\begin{equation}
\frac{d}{dt} \rho = \mathcal{L}_{S} \left [ \rho \right ] = \sum_{n, m} \Gamma^{S}_{nm} \left ( A_{n} \rho A_{m}^{\dagger} - \frac{1}{2} \left \{ A_{m}^{\dagger} A_{n}, \rho \right \}  \right ),
\end{equation}
in which the parameter $t_{\text{op}}$ is inherently distinct from the integration parameter $t$. This noisy dynamics is applied after the coherent dynamics, in order to achieve a first-order approximation to the full Lindblad dynamics,
\begin{equation}
\rho_{0} ~\to~  \rho \left ( t_{\text{op}} \right ) ~=~ \Phi \left [ \rho_{0} \right ] ~\approx~ \exp \left ( t_{\text{op}}\mathcal{L}_{S} \right ) \phi \left [ \rho_{0} \right ].
\end{equation}
We note that the separated noise term is given by a simple exponential, owing to its time-independent nature. Because the rate matrix is time-independent in this sense, we know that the time evolution under the separated noise dynamics will correspond to physical evolution of the density matrix if and only if $\Gamma^{S}$ is PSD. This will be the case whenever the object
\begin{equation}
\Gamma^{F} = \frac{1}{t_{\text{op}}} \int_{0}^{t_{\text{op}}} M^{\dagger} \left ( s \right ) \Gamma^{D}\left ( s \right )  M \left ( s \right )  ds
\end{equation}
is itself a PSD matrix (we refer to this object as $\Gamma^{F}$ because it would be the rate matrix of the ``forward'' noise, which is the form the separated noise would take if it came before, rather than after, the coherent evolution). In particular, this includes the case in which the matrix $\Gamma^{D}$ is PSD at all times.  Furthermore, the total noise strength of $\Gamma^{S}$ will be given by the time-average of the total noise strength of $\Gamma^{D}$,
\begin{equation}
\overline{\gamma}^{S}  = \frac{1}{t_{\text{op}}} \int_{0}^{t_{\text{op}}} \overline{\gamma}^{D} \left ( s \right )  ds .
\end{equation}
In the special case that $\overline{\gamma}^{D}$ is constant, the two noise strengths are precisely equal. We discuss these matters in more detail in Appendix~\ref{sec:appPhys}.

%% file: Sections/justifGiven.tex
\section{Justification for the Gate Noise Mo\-del Given the Separability Ansatz}
\label{sec:justifGiven}

Having stated our main results concerning the separability ansatz, we now argue for the validity of our previously stated model of gate noise. In particular, we argue for the claim that the action of a noisy gate on a quantum device can be modeled as
\begin{equation}
\mathcal{G} \approx \mathcal{N}_{G} G,
\end{equation}
where $\mathcal{G}$ is a quantum operation which results from the application of an imperfect gate, $G$ represents the desired unitary gate operation which would ideally be implemented in the absence of noise, and $\mathcal{N}_{G}$ is a discrete decoherence event. In particular, to lowest order in the noise strength, we can always write
\begin{equation}
\mathcal{N}_{G} \to \exp \left ( t_{G} \mathcal{L}^{G}_{N} \right ),
\end{equation}
where $t_{G}$ is the time required to implement the gate, and $\mathcal{L}^{G}_{N}$ is a term of pure Lindblad form, with no coherent part.

To argue for such a model, we must make some assumptions regarding the nature in which quantum gates are implemented in terms of quantum operations applied to the device, and how these operations are affected by noise. Our fundamental assumption will be that each quantum gate can be expressed as a sequence of individual operations on the qubit degrees of freedom, each of which is Markovian,
\begin{equation}
\mathcal{G}  = \prod_{i} \Phi^{(i)}.
\end{equation}
Each of these operations is generated by some Lindblad generator $\mathcal{L}^{(i)}$ which may in general be time-dependent, defined on the domain $t \in \left (0, t_{\text{op}}^{(i)} \right )$. In other words, we assume that we can model every gate as a sequence of Hamiltonian operations applied to the qubit degrees of freedom, which may occur while the qubits are coupled to an external bath described by Lindblad dissipation. We assume that none of these operations induce any leakage on the device. That is to say, we assume that the interaction with the environment does not map computational states to other states of the device, and thus the number of operational qubits is fixed. We note that the gate which one might hope to implement in the absence of noise is given according to
\begin{equation}
G  = \prod_{i} \phi^{(i)}.
\end{equation}

As an example of such an operation, one may consider a single-qubit rotation around the X-axis, subject to dephasing noise,
\begin{equation}
H = J \sigma^{x} ~;~ \mathcal{L}_{D} \left [ \rho \right ] = \frac{1}{2}\gamma_{\text{deph}} \left[\sigma^{z} \rho \sigma^{z} - \rho\right],
\end{equation}
with some energy scale $J$ and dephasing rate $\gamma_{\text{deph}} $ (the conventional factor of one half in the definition of dephasing noise implies that the quantity $\gamma_{\text{deph}} $ is not precisely the same as the corresponding eigenvalue of the rate matrix). We will assume in all cases that the effects of noise are small compared with those of the coherent time evolution, in the sense that the amount of decoherence which occurs over the characteristic time scale of the Hamiltonian is not too large. For the example above, this would correspond to $\gamma_{\text{deph}} \ll J$. We note that such an assumption is a reasonable requirement to impose upon any NISQ device which is being used to implement a quantum circuit with more than a handful of gates, since otherwise the effects of noise would dominate over any coherent effects to the extent that no meaningful calculation could be performed.

While our fundamental assumptions outlined here will not be valid for all sources of error on all forms of quantum hardware~\cite{PhysRevLett.116.020501, PhysRevLett.121.090502, PhysRevLett.123.190502, PhysRevLett.117.060504, RevModPhys.87.1419, Saffman_2016, cardani_reducing_2021}, describing noise on the device via such Lindblad formalism is well established in the community and proven to be reasonably accurate in practice~\cite{lidar}.

With these assumptions in place, we now argue for the validity of our model of gate noise in two ways. For the first argument, we imagine the quantum gate as consisting of one large quantum operation, defined piece-wise in terms of its individual operations,
\begin{equation}
\begin{split}
& \mathcal{L} \left ( t \right ) \\ \equiv ~ & \mathcal{L}_{H} \left ( t \right ) + \mathcal{L}_{D} \left ( t \right ) \\ \equiv ~ & \sum_{i} r_{i} \left ( t \right ) \mathcal{L}^{(i)}_{H} \left ( t - \tau^{(i)} \right ) + \sum_{i} r_{i} \left ( t \right ) \mathcal{L}^{(i)}_{D} \left ( t - \tau^{(i)} \right ),
\end{split}
\end{equation}
where
\begin{equation}
\tau^{(i)} \equiv \sum_{j<i} t_{\text{op}}^{(j)} 
\end{equation}
and the rectangular function $r_{i} \left ( t \right )$ is defined to be one during the time in which $\mathcal{L}^{(i)}_{H}$ and $\mathcal{L}^{(i)}_{D}$ are active, and zero otherwise. Since nothing in the derivation of the separated noise approximation requires the generator $\mathcal{L}$ to be smooth as a function of time, and since we assume that the noise strength remains appropriately small, we know that the separated noise approximation
\begin{equation}
\Phi = \mathcal{G} \approx  \exp \left ( t_{G}\mathcal{L}_{S} \right ) \phi
\end{equation}
will apply, where in this case the total time of the operation is the gate time $t_{G}$. The coherent part of the evolution is given as usual according to
\begin{equation}
\phi = \mathcal{T} \left \{ \exp \left (  \int_{0}^{t_{G}}  \mathcal{L}_{H} \left ( t \right )  dt \right ) \right \},
\end{equation}
which, due to the piece-wise definition of $\mathcal{L}_{H}$, decomposes as simply
\begin{equation}
\phi = \prod_{i} \mathcal{T} \left \{ \exp \left (  \int_{0}^{t^{(i)}_{\text{op}}}  \mathcal{L}^{(i)}_{H} \left ( t \right )  dt \right ) \right \}   = \prod_{i} \phi^{(i)} = G
\end{equation}
We thus have
\begin{equation}
\mathcal{G} \approx  \exp \left ( t_{G}\mathcal{L}_{S} \right ) G.
\end{equation}
Identifying the separated noise $\mathcal{L}_{S}$ of the entire quantum operation as the gate noise $\mathcal{L}^{G}_{N}$, we ultimately have our desired result. The form of the gate noise can be determined by applying the standard result from Section~\ref{sec:resSum}. We note that the strength of the gate noise will be given by the time average of the noise strength of $\mathcal{L}_{D}$.

For the second argument, we apply the separability ansatz to each individual quantum operation directly, so that the action of the gate becomes, to lowest order in the noise strength,
\begin{equation}
\mathcal{G} \approx \prod_{i} \exp \left ( t^{(i)}_{\text{op}}\mathcal{L}_{S}^{(i)} \right ) \phi^{(i)}.
\end{equation}
We note that the noise strength of each $\mathcal{L}_{S}^{(i)}$ will be given by the time average of the strength of $\mathcal{L}_{D}^{(i)}$, which should be appropriately small. It is clear that if we were to commute all of the terms involving separated noise to the left, the remaining expression would involve the originally desired clean gate, followed by all the terms which have been commuted to the left. Doing so, we find
\begin{equation}
\mathcal{G} \approx \prod_{i}  \exp \left ( t^{(i)}_{\text{op}}\mathcal{L}_{L}^{(i)} \right )\prod_{i}  \phi^{(i)}
\end{equation}
where $\mathcal{L}_{L}^{(i)}$ is the result of commuting $\mathcal{L}_{S}^{(i)}$ past all of the Hamiltonian terms which were originally to its left. Because commuting a noise term past a coherent term does not alter its noise strength, we know that the strength of $\mathcal{L}_{L}^{(i)}$ will be the same as $\mathcal{L}_{S}^{(i)}$, which should again be appropriately small. Thus, it is valid to combine all of the terms $\mathcal{L}_{L}^{(i)}$ together using the Baker-Campbell-Hausdorff (BCH) expansion to lowest order, resulting in
\begin{equation}
\mathcal{G} \approx \exp \left ( \sum_{i} t_{\text{op}}^{(i)}\mathcal{L}_{L}^{(i)} \right )\prod_{i}  \phi^{(i)}.
\end{equation}
If we now associate,
\begin{equation}
t_{G} \mathcal{L}^{G}_{N} ~ \equiv ~ \sum_{i} t^{(i)}_{\text{op}}\mathcal{L}_{L}^{(i)} ~;~ t_{G} ~ \equiv ~ \sum_{i} t^{(i)}_{\text{op}}
\end{equation}
then we ultimately find
\begin{equation}
\mathcal{G} \approx \mathcal{N}_{G} G,
\end{equation}
which is again the desired result. Because we have
\begin{equation}
\sum_{i} \frac{t^{(i)}_{\text{op}}}{t_{G}} = 1,
\end{equation}
then the total noise strength of $\mathcal{L}^{G}_{N}$ is again given by the time average of the total noise strengths of the original noise processes occurring during the implementation of the quantum gate. We note that this second argument is essentially a discrete version of the derivation of the separability ansatz itself, which can be found in Appendix~\ref{sec:appSepDeriv}.

We note that our model of a quantum gate can account for the case in which there is idling time in between individual operations, by modeling this idling time as reflecting the application of the trivial gate, in which no coherent operations are applied to the device. While it is technically true that such an operation would violate our assumption that the effects of decoherence are always smaller than the effects of the coherent evolution, this violation does not ultimately affect the validity of our argument, since during such a trivial operation, the form of the quantum operation is already pure Lindblad, and so the separated noise approximation is already exact, with the separated noise being merely equal to the original noise occurring on the hardware.

%% file: Sections/sepNoiseAcc.tex
\section{Evaluating the Accuracy of the Separated Noise Approximation}
\label{sec:sepNoiseAcc}

Having derived the form of the separated noise, we would like to verify that our expressions are correct, and that they provide a reasonable approximation to the true dynamics of a quantum operation. To this end, we numerically evaluate the time evolution of several systems under the influence of Lindblad dynamics, and then compare these results with a time evolution performed using our approximation, in which the unitary evolution is performed, followed by evolution under the dynamics of the separated noise. We have tested our separated noise formula for several choices of Hamiltonian, underlying hardware noise, observable, and initial state, using both the general, time-dependent formula, as well as the transformation in the special case of time-independent Hamiltonian and noise. We have tested single-qubit systems, as well as multi-qubit systems. In all such cases, we have found exceptional agreement, so long as the noise strength is appropriately small. Here, for simplicity, we will display the results of our analysis for a single-qubit system.

In particular, we consider a single qubit evolving under the Hamiltonian
\begin{equation}
H \left ( t \right ) = \cos \left( t \right ) \sigma^{x} + \sin \left( t \right ) \sigma^{y}.
\end{equation}
We note that this Hamiltonian does not commute with itself at different times, requiring the use of the time-dependent formula in the most general case. In the usual basis of Pauli operators, we find for this Hamiltonian
\begin{equation}
\Omega \left ( t \right ) ~=~
\begin{bmatrix}
   0 & 0 & +2i \sin \left( t \right ) \\
   0 & 0 & -2i \cos \left( t \right ) \\
   -2i \sin \left( t \right ) & +2i \cos \left( t \right ) & 0
   \end{bmatrix}.
\end{equation}
In addition to the coherent dynamics, we assume that the qubit is subject to dephasing noise, in which the dephasing axis oscillates between the Y and Z directions over time,
\begin{equation}
\begin{split}
\mathcal{L}_{D} \left [ \rho \right ] ~ = ~ & \frac{1}{2} \gamma_{\text{deph}} \sin^{2} \left ( \sqrt{2} t \right )  \left [ \sigma^{y} \rho \sigma^{y} - \rho \right ] \\ + ~ & \frac{1}{2} \gamma_{\text{deph}} \cos^{2} \left ( \sqrt{2} t \right )  \left [ \sigma^{z} \rho \sigma^{z} - \rho \right ].
\end{split}
\end{equation}
This corresponds to the rate matrix
\begin{equation}
\Gamma^D ~=~ \frac{1}{2}\gamma_{\text{deph}}
\begin{bmatrix}
   0 & 0 & 0 \\
   0 & \sin^{2} \left ( \sqrt{2} t \right ) & 0 \\
   0 & 0 & \cos^{2} \left ( \sqrt{2} t \right )
   \end{bmatrix}.
\end{equation}
We note that the frequencies of the coherent and incoherent dynamics are incommensurate. We also emphasize that the rate matrix remains PSD at all times.

\begin{figure*}
   \centering
   \includegraphics[width=0.32\textwidth]{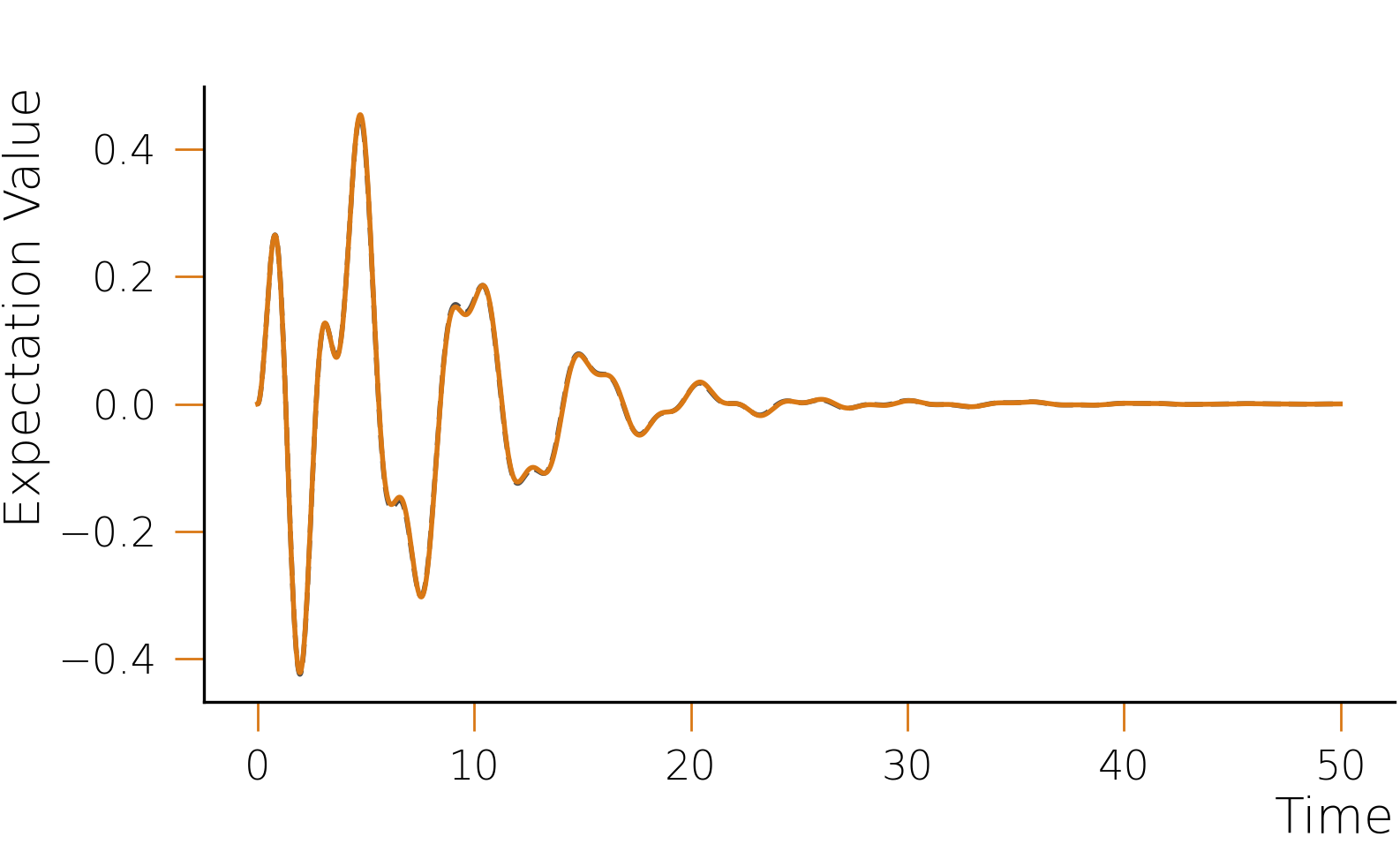}
   \includegraphics[width=0.32\textwidth]{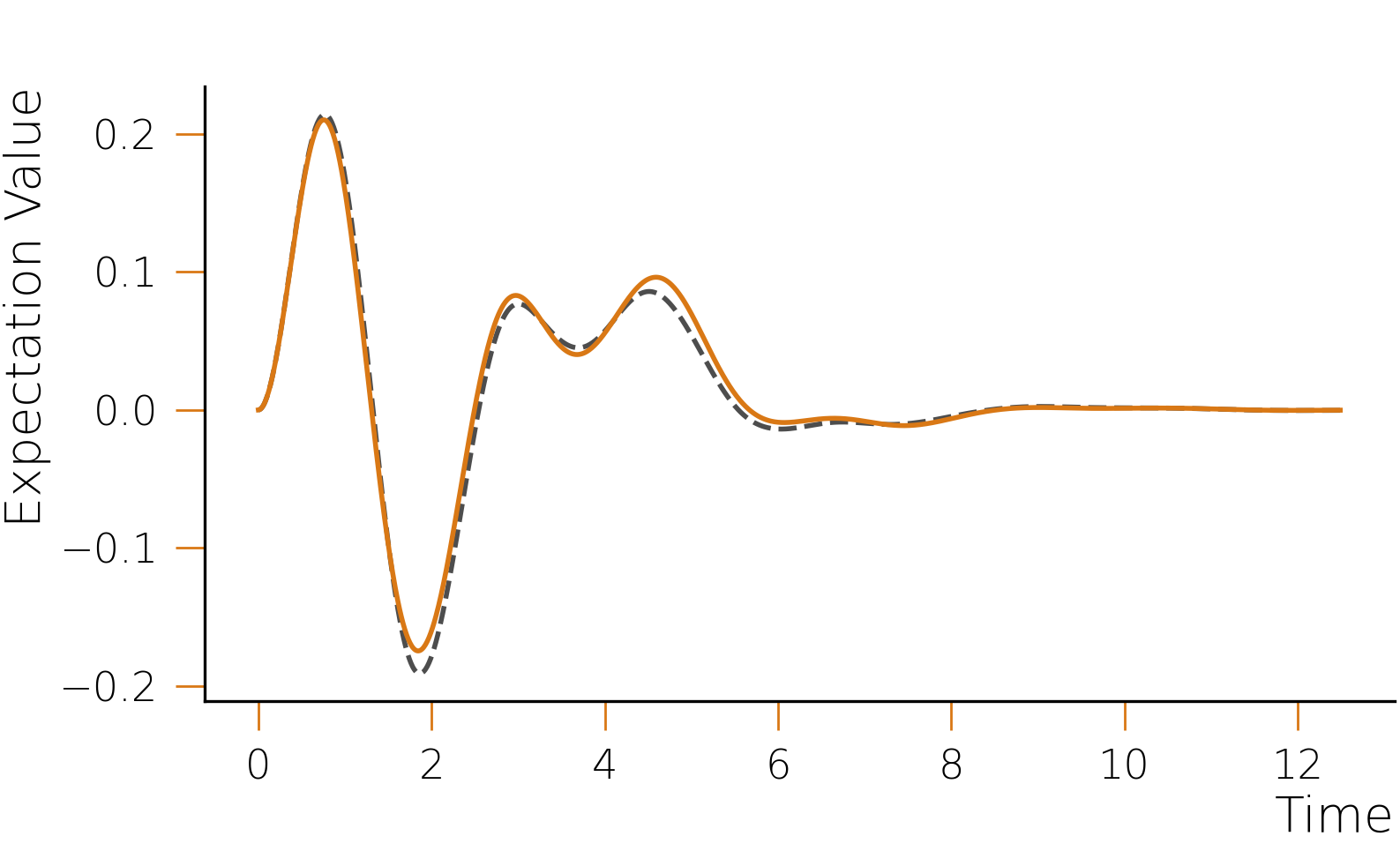}
   \includegraphics[width=0.32\textwidth]{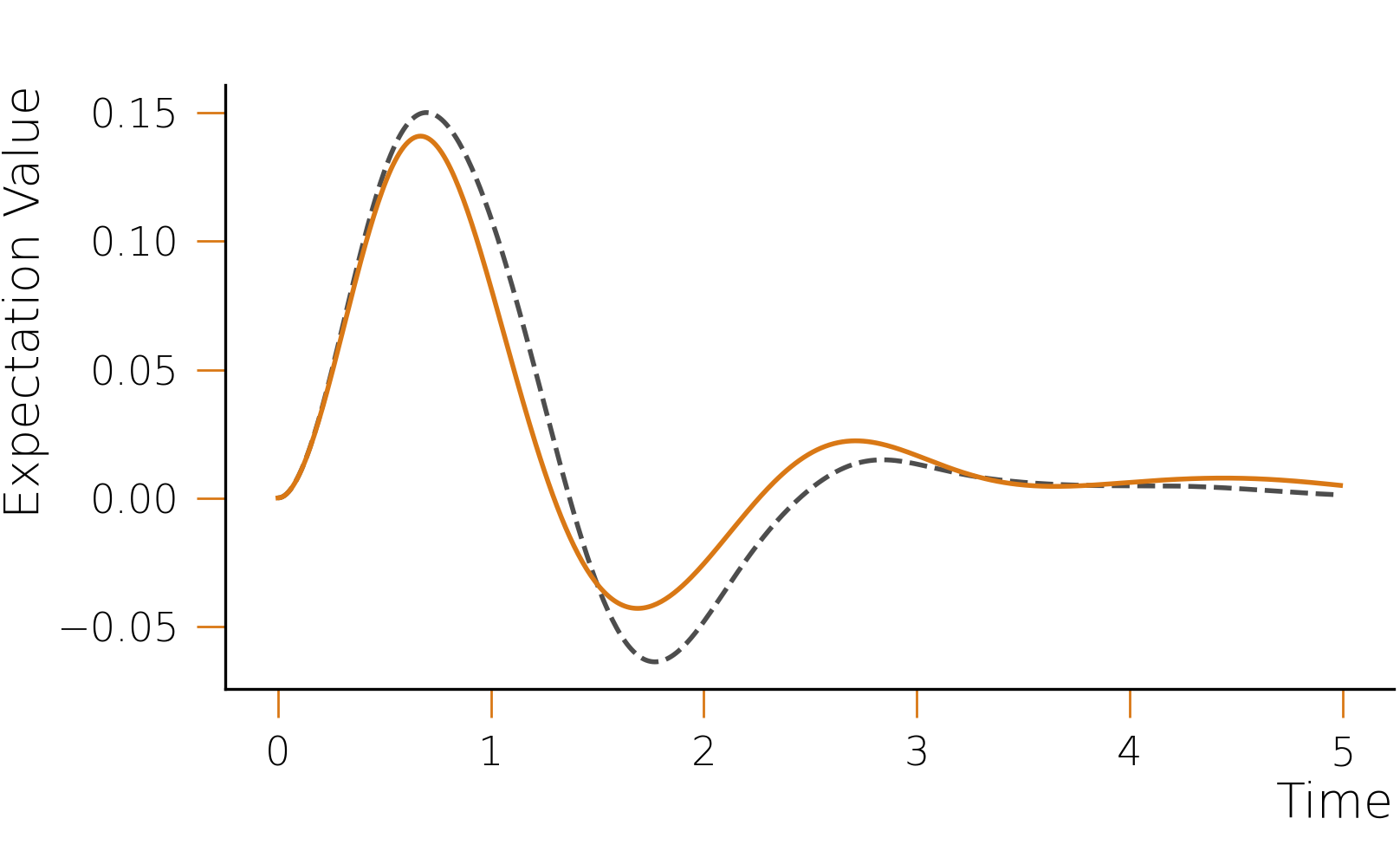}
   \caption{A numerical evaluation of the accuracy of our separated noise approximation, for the case of a single qubit. We compare the separated noise dynamics (dashed line) against the exact dynamics (solid line). The noise strengths, from left to right, are $\gamma_{\text{deph}}$ = 0.25, $\gamma_{\text{deph}}$ = 1.0, and $\gamma_{\text{deph}}$ = 2.5. We see exceptional agreement for weak noise, while the quality of the approximation decreases with increasing noise strength. Note the difference in the time scales of the plots, as the decay time of the system becomes shorter with increasing noise strength.}
   \label{fig:compNoise}
\end{figure*}

In order to find the separated noise, we must solve the differential equation
\begin{equation}
\frac{dQ}{ds} =  -i\left [ \Omega, Q \right ] + \Gamma^{D} ~;~ Q\left ( s = 0 \right ) = 0,
\end{equation}
Solving this differential equation numerically, we can evaluate the accuracy of our approximation for any given operation time. We assume that the spin is fully polarized along the positive Z-axis at time zero. For each subsequent operation time of interest, we first evolve the system under the original combined coherent and noisy dynamics. Then, we perform an evolution under the separated noise approximation, in which we first apply the coherent dynamics, and then the separated noise dynamics. To evaluate the quality of our noise approximation, we will consider the expectation value of $\sigma^{x}$,
\begin{equation}
\langle \sigma^{x}  \rangle = \text{Tr} \left [ \sigma^{x}\rho \left ( t_{\text{op}} \right ) \right ].
\end{equation}
While we choose to display this expectation value, we find similar agreement for the expectation values of $\sigma^{y}$ and $\sigma^{z}$ (and therefore the full density matrix of the single qubit system).

The results of this analysis are shown in Figure~\ref{fig:compNoise}. We note that for small noise strengths, our approximation is exceptionally good, while the quality of the approximation decreases with increasing noise strength. This is in fact exactly what we would expect, as we have derived our results to first order in the noise strength. We note that when the separated noise is a good approximation, it remains that way over the entire evolution of the system, from time zero up until the eventual decay to equilibrium at late times. This is because the separated noise is an approximation which requires the noise strength to be appropriately small in the sense that the effects of noise are not too large over the characteristic time scale of the Hamiltonian. We do not, however, require the effects of noise to be small over the course of the entire time evolution.

%% file: Sections/sepNoiseExamp.tex
\section{Some Examples of Separated Noise}
\label{sec:sepNoiseExamp}

Having given the form of the separated noise and having analyzed its accuracy, we now give some concrete examples of separated noise terms. Here we will focus simply on the single-qubit, time-independent case, since in this case it is possible to evaluate the transformation of the noise explicitly in closed form. This will allow us to provide some additional insight into the nature of the separated noise transformation.

\subsection{Specific Examples of Separated Noise}

Let us consider a general operation with
\begin{equation}
\mathcal{L} \left [ \rho \right ] = -i \left [ H, \rho \right ] + \sum_{\alpha, \beta} \Gamma^{D}_{\alpha \beta} \left [ \sigma^{\alpha} \rho \sigma^{\beta} - \frac{1}{2} \left \{ \sigma^{\beta} \sigma^{\alpha}, \rho \right \} \right ],
\end{equation}
where the $\left\{\sigma^{\alpha}\right\}$ are the usual Pauli operators acting on a single qubit. The rate matrix of the noise term will in general be a $3\times3$ positive semi-definite matrix, 
\begin{equation}
\Gamma^{D} = 
\begin{bmatrix}
   \Gamma^{D}_{XX} & \Gamma^{D}_{XY} & \Gamma^{D}_{XZ} \\
   \Gamma^{D}_{YX} & \Gamma^{D}_{YY} & \Gamma^{D}_{YZ} \\
   \Gamma^{D}_{ZX} & \Gamma^{D}_{ZY} & \Gamma^{D}_{ZZ}
   \end{bmatrix}.
\end{equation}
As for the Hamiltonian term, we choose to take
\begin{equation}
H = - J \sigma^{x},
\end{equation}
for some $J > 0$. Any other choice of Hamiltonian is equivalent through rotational symmetry. We will be interested in understanding how the underlying hardware noise transforms into separated noise under this particular choice of Hamiltonian, for various choices of hardware noise. Since we consider the time-independent case, we can use the alternate expression
\begin{equation}
\Gamma^{S} = \mathcal{K} \left [ \Gamma^{D} \right] = \sum_{m=0}^{\infty} \frac{ t_{\text{op}}^{m}}{(m+1)!} \xi^{m} \left [ \Gamma^{D} \right ],
\end{equation}
which we find to be somewhat more useful than the differential equation (\ref{eqn:mainDiff}) in the analysis that follows. In Appendix~\ref{sec:appNumerics}, we discuss how to evaluate the above transformation. Here, we simply quote the results of our analysis. We will display our results in terms of the operation angle, rather than operation time, defined as
\begin{equation}
\theta \equiv 2 J t_{\text{op}}.
\end{equation}

\begin{figure}
   \centering
   \includegraphics[width=0.5\textwidth]{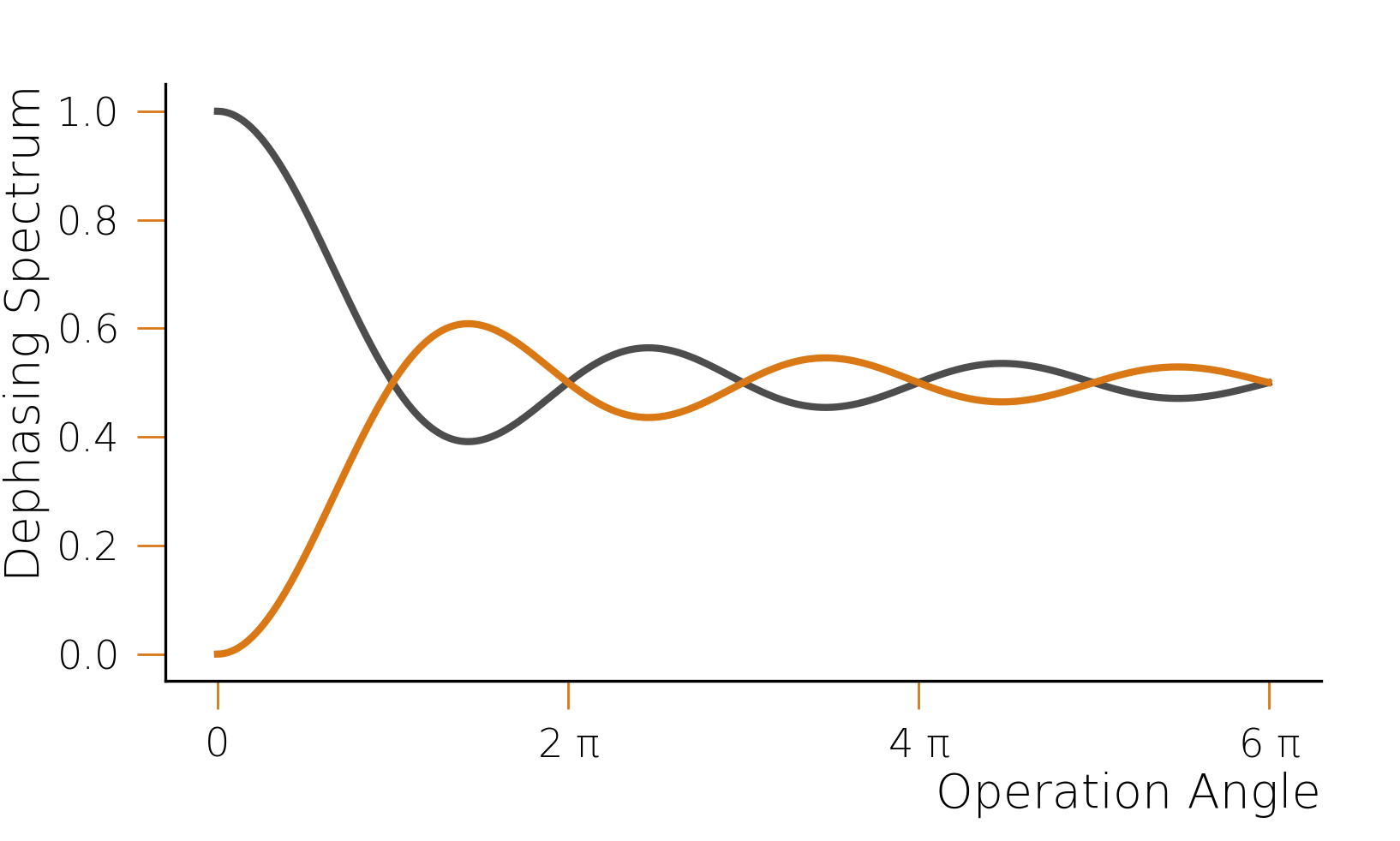}
   \caption{The transformed dephasing spectrum, as a function of operation angle, in units of $\gamma_{\text{deph}}/2$.}
   \label{fig:deph}
\end{figure}

Beginning with the case of dephasing noise on a single qubit, the rate matrix is
\begin{equation}
\Gamma^{D} = \frac{1}{2}\gamma_{\text{deph}}
\begin{bmatrix}
   0 & 0 & 0 \\
   0 & 0 & 0 \\
   0 & 0 & 1
   \end{bmatrix},
\end{equation}
where $\gamma_{\text{deph}}$ is the dephasing rate. Applying the transformation to this matrix, we find
\begin{equation}
\Gamma^S = \frac{\gamma_{\text{deph}}}{4\theta}
\begin{bmatrix}
   0 & 0 & 0 \\
   0 & \theta - \frac{1}{2}\sin(2\theta) & \sin^2(\theta) \\
   0 & \sin^2(\theta) & \theta + \frac{1}{2}\sin(2\theta)
\end{bmatrix}.
\end{equation}
Note how the trace and Hermitian nature of the rate matrix are explicitly preserved. It is clear that for general values of the parameter $\theta$, the noise can deviate quite significantly in nature from its original form, regardless of how the operation time compares with the dephasing rate $\gamma_{\text{deph}}$. The spectrum of the separated noise can be found from the non-zero eigenvalues of the above matrix, of which there are two,
\begin{equation}
\gamma_{\pm} = \frac{1}{4} \gamma_{\text{deph}} \left ( 1 \pm \frac{\sin (\theta)}{\theta} \right ).
\end{equation}
A plot of these eigenvalues as a function of the parameter $\theta$ is shown in Figure~\ref{fig:deph}. We see explicitly how the positivity of the spectrum is preserved by the transformation.

\begin{figure}
   \centering
   \includegraphics[width=0.5\textwidth]{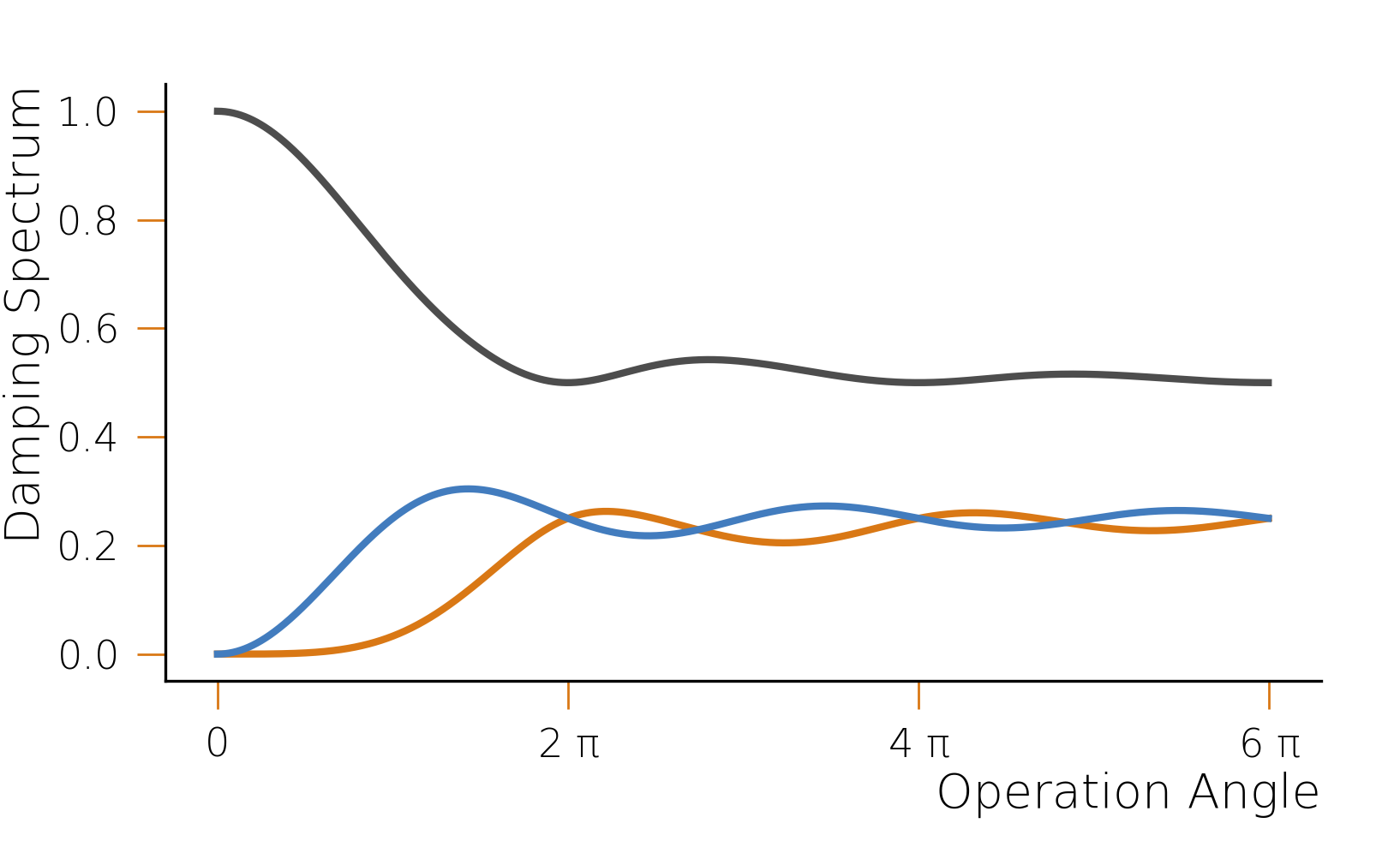}
   \caption{The transformed damping spectrum, as a function of operation angle, in units of $\gamma_{\text{damp}}/2$.}
   \label{fig:damp}
\end{figure}

Another important example of single qubit noise is damping noise, given by
\begin{equation}
\Gamma^{D} = \frac{\gamma_{\text{damp}}}{4}
\begin{bmatrix}
   1 & -i & 0 \\
   +i & 1 & 0 \\
   0 & 0 & 0
   \end{bmatrix}.
\end{equation}
In this case, the transformed rate matrix is somewhat more complicated,
\begin{equation}
\begin{split}
& \Gamma^S =
\frac{\gamma_{\text{damp}}}{8\theta} \times \\
& \begin{bmatrix}
 2\theta & -2i \sin (\theta) & -2i \left (\cos (\theta) - 1 \right ) \\
 2i \sin (\theta) & \theta + \frac{1}{2} \sin(2\theta) & -\sin ^2(\theta) \\
 2i \left ( \cos (\theta) - 1 \right ) & -\sin ^2(\theta) & \theta - \frac{1}{2} \sin(2\theta)\\
\end{bmatrix}.
\end{split}
\end{equation}
Likewise, the spectrum of this rate matrix is now more complicated, with three non-zero noise rates given by
\begin{equation}
\begin{split}
\gamma_{0} &= \frac{1}{8} \gamma_{\text{damp}} \left (1 -  \frac{\sin (\theta)}{\theta} \right ) ~;~ \\
\gamma_{\pm} &= \frac{3}{16}\gamma_{\text{damp}} + \frac{1}{16\theta}\gamma_{\text{damp}} \Big[ \sin (\theta) \\ & \ \  \pm  \sqrt{\left(\theta - \sin (\theta)\right)^2 + 32 \left(1 - \cos (\theta)\right)} \Big].
\end{split}
\end{equation}
Again, a plot of this spectrum is shown in Figure~\ref{fig:damp}.

Lastly, we consider depolarizing noise acting on a single qubit. In this case, the rate matrix is proportional to the identity matrix, and we in fact have
\begin{equation}
\xi \left [\Gamma^{D} \right ] = -i \left [\Omega, \Gamma^{D} \right ] = 0 ~ \Rightarrow ~ \mathcal{K} \left [ \Gamma^{D} \right ] = \Gamma^{D}.
\end{equation}
Thus, depolarizing noise possesses the property that it can be separated without any change to its structure. This is merely a reflection of the fact that single-qubit depolarizing noise commutes with any single-qubit coherent dynamics, and thus separating the exponential terms in this case is exact. We emphasize, however, that this only holds for the case of one qubit. For multiple qubits subject to independent, single-qubit depolarizing noise, the rate matrix is not simply proportional to the identity matrix. The case in which the rate matrix is indeed proportional to the identity matrix corresponds instead to applying a depolarizing channel to the global density matrix of the entire system,
\begin{equation}
\rho \to (1-\lambda)\rho+\frac{\lambda}{\mathcal{D}}I,
\end{equation}
with $\lambda$ related to the proportionality constant $\alpha$ by
\begin{equation}
\lambda = 1 - e^{-\alpha \mathcal{D}^{2} t_{\text{op}}}.
\end{equation}
It is this global depolarizing channel which can therefore be considered separately from the coherent dynamics, in the case of a multi-qubit system.

In the two-qubit case, the rate matrix has dimensions $15 \times 15$, and the structure of the map $\mathcal{K}$ becomes significantly more complicated. We do not attempt an analytic evaluation of the separated noise in this case. However, one important feature of the two-qubit case which we wish to emphasize here, is that the map $\mathcal{K}$ can produce separated noise which is correlated, even when the underlying hardware noise is not correlated.

\subsection{Angular Dependence of the Separated Noise}

In the preceding examples, the separated noise depended on the ``angle'' implemented by the coherent term, rather than the operation time, or its relation to the time scale of the underlying hardware noise. This is indeed a generic feature of the separated noise. If we define a dimensionless, scaled $\overline{\Omega}$ according to,
\begin{equation}
\Omega \equiv 2 J \overline{\Omega}
\end{equation}
which corresponds to dividing the Hamiltonian by its overall energy scale, then our transformation of the rate matrix can alternatively be written as
\begin{equation}
\Gamma^{S} = \frac{1}{ \theta} M  \left [ \int_{0}^{\theta} M^{\dagger} \left ( \varphi \right ) \Gamma^{D} \left ( \varphi \right )  M \left ( \varphi \right ) d\varphi \right ] M^{\dagger},
\end{equation}
where
\begin{equation}
M \left ( \varphi \right ) \equiv \mathcal{T} \left \{ \exp \left ( -i \int_{0}^{\varphi} \overline{\Omega} \left ( \lambda \right ) d\lambda \right )   \right \}.
\end{equation}
In the small angle limit, the matrix $M$ approaches the identity transformation, and we simply have
\begin{equation}
\lim_{\theta \to 0} \Gamma^{S} = \Gamma^{D}.
\end{equation}
We emphasize that we are able to obtain the above limit without making any statements about how the time scale of the coherent dynamics compares with the time scales of the underlying hardware noise. This suggests that the relevant  criterion for when the separated noise deviates substantially from the underlying hardware noise involves the magnitude of the angle implemented by the coherent evolution, rather than any considerations regarding how ``quickly'' the coherent dynamics occur.

\subsection{Steady State Behavior of the Separated Noise}

One striking feature of the separated noise examples presented here is that their spectrum appears to approach a steady state with increasing operation angle, thus suggesting that the separated noise itself reaches a steady state value. This is in fact also a generic feature of the separated noise, at least in the time-independent case. In particular, if we consider the underlying rate matrix $\Gamma^{D}$ as a vector in the space of Hermitian matrices, and $\xi$ to be an operator on this space, then the steady state value of $\Gamma^{S}$ is given by the projection of $\Gamma^{D}$ onto the null space of $\xi$. 

We prove the above claim in Appendix~\ref{sec:appSteady}. To see a concrete example, we consider again the case of dephasing noise from the beginning of this section. For our chosen single-qubit Hamiltonian,
\begin{equation}
\Omega ~=~
\begin{bmatrix}
   0 & 0 & 0 \\
   0 & 0 & +2Ji  \\
   0 & -2Ji & 0
   \end{bmatrix}
\end{equation}
This object is in fact proportional to the seventh Gell-Mann matrix,
\begin{equation}
\Omega ~=~ -2J \lambda_{7}.
\end{equation}
We now must find the null space of the operator
\begin{equation}
\xi = -i \text{ad}_{\Omega} = +2 J i \left [ \lambda_{7}, \cdot \right ].
\end{equation}
This amounts to finding the space of all Hermitian $3 \times 3$ matrices which commute with $\lambda_{7}$.

The space of all Hermitian three by three matrices can be decomposed into the space of all traceless, Hermitian $3 \times 3$ matrices (in other words, the lie algebra $\mathfrak{su}(3)$), together with the identity matrix. Thus, to form a basis for all Hermitian $3 \times 3$ matrices, we can consider the Gell-Mann matrices, together with the identity. Finding the null space of $\xi$ can be accomplished by evaluating its matrix elements in this basis, and then diagonalizing the resulting matrix, isolating the eigenvectors with zero eigenvalue as forming a basis of the null space. Doing so, we find that the null space is given by
\begin{equation}
\text{null} \left ( \xi \right ) = \text{span} \left \{ \frac{1}{\sqrt{3}}I, \frac{1}{\sqrt{2}} \lambda_{7},\frac{1}{2\sqrt{2}} \left ( \sqrt{3}\lambda_{3} + \lambda_{8}  \right )  \right \},
\end{equation}
where we have chosen to normalize this basis of rate matrices with respect to the usual Frobenius inner product, in this case \textbf{without} any factors involving the Hilbert space dimension $\mathcal{D}$.

The steady state value of $\Gamma^{S}$ is given by the projection of 
\begin{equation}
\Gamma^{D} = \frac{1}{2}\gamma_{\text{deph}}
\begin{bmatrix}
   0 & 0 & 0 \\
   0 & 0 & 0 \\
   0 & 0 & 1
   \end{bmatrix}
\end{equation}
into this null space. Performing this projection, we find
\begin{equation}
\Gamma^{S}_{\infty} = \frac{1}{2}\gamma_{\text{deph}}
\begin{bmatrix}
   0 & 0 & 0 \\
   0 & \frac{1}{2} & 0 \\
   0 & 0 & \frac{1}{2},
   \end{bmatrix}
\end{equation}
which is indeed the steady state value of the expression we previously found.

In regards to the angular scale over which the separated noise approaches its steady state value, we note two things. First, we can see from the explicit form of the separated noise that this angular scale does not depend on either the energy scale of the Hamiltonian, or the overall strength of the original noise. Second, the separated noise appears to periodically reach its steady state value already at finite values of the operation angle. We explore both of these phenomena in more detail in Appendix~\ref{sec:appSteady}.

\subsection{An Example of Gate Noise}

Before closing this section, we briefly give an example of a noise term following a gate which is composed of multiple quantum operations, therefore connecting our discussion of separated noise back to the original motivation of our work. In particular, we assume a toy model of an ideal Hadamard gate, whose unitary time-evolution matrix in the noise-free case would correspond to
\begin{equation}
U = R_y(\pi/2)Z = \exp(-i\frac{\pi}{4}Y)\exp(-i\frac{\pi}{2}(I-Z)).
\end{equation}
Since the global phase introduced by this single-qubit gate is not measurable in practice, we study simply
\begin{equation}
U \to \exp(-i\frac{\pi}{4}Y)\exp(+i\frac{\pi}{2}Z).
\end{equation}
Our toy model thus corresponds to a sequence of two elementary rotations. In addition to the coherent evolution, we assume that there is hardware noise occurring during the implementation of this gate, which in this case we take to be damping noise,
\begin{equation}
\mathcal{L}_{\text{damp}} \left [ \rho \right ] =  \gamma_{\text{damp}}  \left [ \sigma^{+} \rho \sigma^{-} - \frac{1}{2} \left \{\sigma^{-}\sigma^{+}, \rho \right \}  \right ]
\end{equation}
We wish to model the effects of this noise in the usual way,
\begin{equation}
\mathcal{G} \approx \mathcal{N}_{G} G,
\end{equation}
where $\mathcal{G}$ is the gate under the influence of noise, $G$ is the action of the ideal gate,
\begin{equation}
G \to U \left ( \cdot \right ) U^{\dagger},
\end{equation}
and $\mathcal{N}_{G}$ is the discrete noise term
\begin{equation}
\mathcal{N}_{G} = \exp \left ( t_{G} \mathcal{L}^{G}_{N} \right ).
\end{equation}

To accomplish this, we must compute the separated noise term for each elementary rotation, commute the separated terms to the left, and add them together. We discuss this process in more detail in Appendix~\ref{sec:appGateExamp}. Here, we simply quote the result. If we make the assumption that the first elementary rotation around the Z axis takes twice as long to implement as the second elementary rotation around the Y axis, then we ultimately find the rate matrix of the gate noise to be
\begin{equation}
\Gamma^{N} ~=~
\frac{1}{2}\gamma_{\text{damp}}
\begin{bmatrix}
\frac{1}{12} & -\frac{i}{3 \pi } & -\frac{1}{6 \pi } \\
 \frac{i}{3 \pi } & \frac{1}{2} & -\frac{i (1+\pi )}{3 \pi } \\
 -\frac{1}{6 \pi } & \frac{i (1+\pi )}{3 \pi } & \frac{5}{12} \\
\end{bmatrix}
\end{equation}
or, evaluated numerically,
\begin{equation}
\Gamma^{N} \approx
\frac{1}{2}\gamma_{\text{damp}}
\begin{bmatrix}
0.083 & -0.106 i & -0.053 \\
0.106i & 0.5 & -0.439 i \\
 -0.053 & 0.439 i & 0.417 \\
\end{bmatrix}.
\end{equation}
Despite the relatively simple nature of both the underlying hardware noise and the coherent gate, the effective noise term can take a relatively complex form. We emphasize that we have arrived at this result without making any assumptions about the magnitude of $\gamma_{\text{damp}}$, or its relationship to the operation times $t^{Z}_{\text{op}}$ and $t^{Y}_{\text{op}}$ of the $Z$ and $Y$ rotations.

The above model of gate noise assumes that the two elementary operations are performed immediately one after another, without any significant idling time between them. If this assumption is violated, then the specific form of the gate noise can change. In particular, in the limit that the idling time is large compared with the duration of each elementary operation, the gate noise will simply approach the underlying hardware noise. We emphasize, however, that this approach to the hardware noise depends only on the ratio of the idling time to the operation time. The time scale of the noise itself is not relevant. This is discussed in more detail in Appendix~\ref{sec:appGateExamp}.

%% file: Sections/applications.tex
\section{Applications}
\label{sec:applications}

We briefly discuss here some potential applications of the results we have found in this work. First and foremost, our results justify the use of the discrete noise approximation when modeling the effects of noise on quantum circuits, for all cases in which Lindblad-like noise is a reasonable description of the underlying decoherence processes occurring on the quantum device. In particular, we use this justification when modeling the effects of noise in our other works \cite{fratus2022describing, leppaekangas2023quantum}, in which our analysis of the effects of noise at the circuit level is greatly simplified through the use of the discrete noise approximation.

Beyond this immediate application of our work, there are several other potential applications, which we discuss below. Some of these are more speculative in nature, therefore motivating additional research.

First, while we have provided a justification for the discrete noise approximation, and considered several examples of separated noise, we have not said much about how these results may be useful in providing the user of a quantum device with additional control over the circuit which is being implemented. We have seen how the form of the separated noise depends on the magnitude of the rotation implemented by the Hamiltonian, and how this separated noise determines the circuit-level description of quantum gates followed by discrete noise operations. This implies that the form of the separated noise, and thus the circuit-level noise, may be modified by judicious choice of Hamiltonian terms and the magnitude of their associated rotations. Our results provide a quantitative description of the separated noise which can be used when implementing such a protocol, thereby complementing existing methods in the field of noise-tuning \cite{Sun2021}.

Second, we have not explored the question of separated noise steady states in much detail, and have thus far said nothing about their potential application beyond the current work. If it is possible to isolate the steady state behavior of the separated noise for a given Lindblad dynamics, and if this separated noise takes a reasonably simple form, this may assist in the numerical simulation of said Lindblad dynamics. This may be the case, for example, if the separated noise steady state represents a noisy dynamics which is much more computationally tractable to simulate than the original full dynamics. In such a case, one could perform the unitary dynamics, which can be implemented on the much smaller space of pure states, followed by the separated noise dynamics. This could prove especially useful in cases where the separated noise reaches its steady state value before the quantum system has reached equilibrium under the effects of noise, as the dynamics being simulated would still be out of equilibrium, and thus non-trivial. Naively, this may seem like a significantly more challenging problem than simply solving the original Lindblad dynamics, since solving the differential equation which defines the separated noise involves simulating the ``dynamics'' of an object which has dimensions $\left ( \mathcal{D}^{2} - 1 \right ) \times \left ( \mathcal{D}^{2} - 1 \right )$, rather than the dynamics of the density matrix, which has dimensions $\mathcal{D} \times \mathcal{D}$. However, since the transformation $\xi$ which defines this differential equation corresponds to the adjoint action of the adjoint action of the Hamiltonian, it is possible to construct the eigenmodes of $\xi$ directly from the eigenstates of the Hamiltonian, as described in Appendix~\ref{sec:appNumerics}. These eigenmodes can be used to find the steady state of the separated noise, as discussed in Appendix~\ref{sec:appSteady}. Additionally, there exist approximate numerical methods \cite{Kuo:22} for projecting vectors into the null space of linear operators, which may also be useful in finding the steady state separated noise. It is not immediately clear which techniques may be available in the fully time-dependent case, thus further motivating the need for further research into this subject.

A third potential application of our results could be to the method of zero-noise extrapolation \cite{PhysRevX.7.021050, PhysRevLett.119.180509}. In some cases, simulation of a noisy system may be simpler to implement numerically, as opposed to the corresponding clean dynamics (due to, perhaps, the ability to make use of a restricted state space representation when the spread of correlations in the system is limited due to noise). It may also be the case that we have effectively simulated a noisy quantum system by simulating an otherwise noise-free system on a noisy quantum computer \cite{fratus2022describing}, and we have an understanding of the effective noise model being simulated, due to an ability to adequately characterize the noise processes occurring on the hardware. In both such cases, we could imagine inverting the relationship which defines the separated noise, in order to find
\begin{equation}
\phi = \exp \left ( - t_{\text{op}}\mathcal{L}_{S} \right )\Phi
\end{equation}
In the event that we are able to evaluate (at least approximately) the form of $\mathcal{L}_{S}$, this could provide a rigorous expression for recovering the clean dynamics from the noisy dynamics, in the limit of weak noise, without needing to make any assumptions about whether we are in the linear or exponential noise regime.

Lastly, we note that our condition for the separated noise dynamics to be physical may actually tell us something (at least in the limit of weak noise) about the physical nature of the Lindblad equation in the general time-dependent case, which currently remains an open question \cite{RevModPhys.88.021002}. To understand why, we recall that the separated noise represents a method for approximating the original full dynamics,
\begin{equation}
\Phi \approx \exp \left ( t_{\text{op}} \mathcal{L}_{S} \right ) \phi.
\end{equation}
Since the unitary dynamics will always be physical, the physical nature of $\Phi$ depends entirely on the physical nature of the separated noise term. Furthermore, we know that the separated noise will be physical if and only if
\begin{equation}
\Gamma^{F} = \frac{1}{t_{\text{op}}} \int_{0}^{t_{\text{op}}} M^{\dagger} \left ( s \right ) \Gamma^{D}\left ( s \right )  M \left ( s \right )  ds
\end{equation}
is a positive semi-definite matrix. This therefore places a constraint on the relationship between the noisy and coherent dynamics in order for the full dynamics to be physical, at least in the limit of weak noise. It may be an interesting problem to consider the extent to which this result could provide additional insight into the physical nature of the Lindblad equation in the general time-dependent case, and whether this result could be extended beyond lowest order in the noise strength.

The potential applications mentioned above indicate that our results presented here could possibly serve as the basis for significant future research.

%% file: Sections/conclusion.tex
\section{Conclusion}
\label{sec:conclusion}

In this work, we have analyzed the validity of the discrete noise approximation in quantum circuits, in which the effects of noise on a quantum circuit are modeled by individual decoherence events following each gate. In the process of doing so, we have found an approximation to the Lindblad dynamics of a general noisy system which is valid to lowest order in the noise strength, in which the unitary dynamics are performed, followed by purely Lindblad dynamics. These considerations have been supplemented with analytic proofs, numerical verification, and concrete examples regarding the manner in which these objects can be computed in practice. Our results generally support the discrete noise approximation, however, they also indicate that in many cases, the form of the effective noise may differ significantly from that of any underlying noise processes occurring on the quantum device. This highlights the importance of accurately modeling noise processes occurring on a quantum device at the circuit level, and may call into question models of circuit noise which invoke relatively simple noise processes (for example, single-qubit damping or dephasing). We have also discussed several potential applications of these results, many of which extend beyond the original scope of this work. We hope that these results presented here will allow for a better understanding of the subject of noise in quantum computation, as well as the subject of Lindblad dynamics more broadly.

%% file: Sections/acknowledgment.tex
\acknowledgments{
This work received funding from the European Union’s Horizon program with numbers 899561 (AVaQus) and 101046968 (BRISQ). We thank Eric Dzienkowski, Sebastian Fischetti, and Brayden Ware for helpful discussions.
}

%% file: Sections/appSum.tex
\section*{Appendix}

The appendices below contain a variety of material which is supplementary to the main text. Here we briefly outline the information contained within them (in somewhat more detail than the description given in the introduction to the main text). In Appendix~\ref{sec:appCommuteGate} we derive various results involving the commutators of coherent and incoherent quantum operations, which we use in the derivation of our main results concerning the separability ansatz. In particular, we show that the effect of commuting a Lindblad noise term past a coherent quantum operation results in another Lindblad noise term, and we provide a quantitative description as to how to compute such an object. We also derive the form of the commutator between a coherent generator $\mathcal{L}_{H}$ and a Lindblad dissipator $\mathcal{L}_{D}$. In Appendix~\ref{sec:appSepDeriv}, we derive the main results concerning the separability ansatz. We first derive the general integral expression in two ways, followed by a derivation of the differential equation involving the transformation $\xi$. We also give an alternate derivation of the separated noise in the time-independent case. In Appendix~\ref{sec:appPhys} we provide an analytic discussion of the necessary conditions for the separated noise term to be physical, and give additional numerical verification of the physical nature of the separated noise in the time-independent case. In Appendix~\ref{sec:appNumerics}, we elaborate in more detail on how the separated noise can be computed in practice in the time-independent case, using both numerical and analytic methods. In Appendix~\ref{sec:appSteady} we derive the form of the separated noise steady state in the time-independent case, and discuss the manner in which the separated noise decays to such a steady state. In Appendix~\ref{sec:appGateExamp} we provide an explicit derivation of a discrete noise term following a full quantum gate, using the methods derived in our work.

%% file: Sections/appCommuteGate.tex
\section{The Effect of Commuting a Lindblad Term Past a Unitary Quantum Operation}
\label{sec:appCommuteGate}

In the process of analyzing the separability ansatz, we have found that it is necessary to understand how a noise term is modified when it is commuted past a unitary gate. More concretely, we would like to solve the following equation,
\begin{equation}
\phi e^{\mathcal{L}_{D}} = e^{\mathcal{L}_{L}} \phi,
\end{equation}
where $\phi$ represents some noise-free quantum operation, $\mathcal{L}_{D}$ describes the original noise term, and $\mathcal{L}_{L}$ describes the modified noise term we wish to find (we will assume for simplicity that the dissipator $\mathcal{L}_{D}$ is dimensionless - that is to say, we will absorb the time under which it acts into the definition of its rate matrix). Since the inverse of a coherent quantum operation always represents another valid coherent evolution, we can always write
\begin{equation}
e^{\mathcal{L}_{L}} = \phi e^{\mathcal{L}_{D}}\phi^{-1}.
\end{equation}
It is thus clear that the term on the left will always correspond to some physical evolution of the system, since it represents a sequence of other physical operations. However, what is less clear is the precise form of this operation. In this appendix, we show that the term $\mathcal{L}_{L}$ corresponds to another Lindblad dissipator, and we provide various expressions for evaluating its precise form.

\subsection{Derivation of the Form of the Commuted Noise}

The equation we would like to solve is
\begin{equation}
\phi e^{\mathcal{L}_{D}} = e^{\mathcal{L}_{L}} \phi,
\end{equation}
where $\phi$ is some coherent quantum operation given according to
\begin{equation}
\phi = \mathcal{T} \left \{ \exp \left ( \int_{0}^{t_{\text{op}}} \mathcal{L}_{H} \left ( t \right ) dt \right ) \right \}.
\end{equation}
Acting on a density matrix, the coherent operation can be written as
\begin{equation}
\phi \left [ \rho \right ] = U \rho U^{\dagger},
\end{equation}
where the unitary matrix $U$ is given according to
\begin{equation}
U = \mathcal{T} \left \{ \exp \left ( -i\int_{0}^{t_{\text{op}}} H \left ( t \right ) dt \right ) \right \}.
\end{equation}
Since $U$ is a unitary matrix, there always exists some hermitian operator $K$ such that
\begin{equation}
U = e^{-it_{\text{op}}K} ~ \Rightarrow ~ \phi \left [ \rho \right ] =  e^{-it_{\text{op}}K}~ \rho ~e^{+it_{\text{op}}K}.
\end{equation}
Using the identity
\begin{equation}
\text{Ad}_{e^{X}} = e^{\text{ad}_{X}},
\end{equation}
where
\begin{equation}
\text{Ad}_{e^{X}} \left ( Y \right ) \equiv e^{X} Y e^{-X},
\end{equation}
we can further write
\begin{equation}
\phi \left [ \rho \right ] ~=~ e^{-it_{\text{op}} \text{ad}_{K}} \rho ~\equiv~ e^{t_{\text{op}}\mathcal{L}_{K}} \rho.
\end{equation}
We note that in the general time-dependent case, $K$ will not necessarily be equal to $H$. Our equation to be solved can thus be written
\begin{equation}
e^{t_{\text{op}}\mathcal{L}_{K}} e^{\mathcal{L}_{D}} = e^{\mathcal{L}_{L}} e^{t_{\text{op}}\mathcal{L}_{K}}.
\end{equation}
The solution to this equation is the braiding identity:
\begin{equation}
\mathcal{L}_{L} = e^{t_{\text{op}}\mathcal{L}_{K}} \mathcal{L}_{D} e^{-t_{\text{op}}\mathcal{L}_{K}} = \phi \mathcal{L}_{D} \phi^{-1}.
\end{equation}

In order to explicitly determine the structure of $\mathcal{L}_{L}$, we first note that
\begin{equation}
U^{\dagger} \left ( U \rho U^{\dagger} \right ) U = \rho ~ \Rightarrow ~ \phi^{-1} \left [ \rho \right ] = U^{\dagger} \rho U.
\end{equation}
We can thus write
\begin{equation}
\mathcal{L}_{L} \left [ \rho \right ] = \left ( \phi \mathcal{L}_{D} \phi^{-1} \right ) \left [ \rho \right ] = U \left ( \mathcal{L}_{D} \left [ U^{\dagger} \rho U \right ] \right ) U^{\dagger}
\end{equation}
Using the explicit form of the Lindblad dissipator,
\begin{equation}
\mathcal{L}_{D} \left [ \rho \right ]  = \sum_{n,m} \Gamma^{D}_{nm}\left[A_n\rho A_m^\dagger-\frac{1}{2}\left\{A_m^\dagger A_n, \rho\right\}\right],
\end{equation}
we have
\begin{equation}
\begin{split}
\mathcal{L}_{L} \left [ \rho \right ] = U \Bigg[ \sum_{n,m} \Gamma^{D}_{nm} & \bigg[ A_n  \left ( U^{\dagger} \rho U \right ) A_m^\dagger \\
& -\frac{1}{2}\left\{ A_m^{\dagger} A_n, \left ( U^{\dagger} \rho U \right ) \right\} \bigg] \Bigg] U^{\dagger}.
\end{split}
\end{equation}
A straight-forward manipulation of this equation brings it into the alternate form,
\begin{equation}
\begin{split}
\mathcal{L}_{L} \left [ \rho \right ] = \sum_{n,m}  \Gamma^{D}_{nm} & \bigg[ \left ( U A{_n} U^{\dagger} \right )  \rho  \left ( U A{_m} U^{\dagger} \right )^{\dagger} \\
& - \frac{1}{2}\left\{\left ( U A{_m} U^{\dagger} \right )^{\dagger} \left ( U A{_n} U^{\dagger} \right ), \rho \right\}\bigg],
\end{split}
\end{equation}
or, alternatively,
\begin{equation}
\begin{split}
\mathcal{L}_{L} \left [ \rho \right ] = \sum_{n,m}  \Gamma^{D}_{nm} & \bigg[ \left ( \phi \left [  A_{n} \right ] \right )  \rho  \left ( \phi \left [  A_{m} \right ] \right )^{\dagger} \\
& - \frac{1}{2}\left\{\left ( \phi \left [  A_{m} \right ] \right )^{\dagger} \left ( \phi \left [  A_{n} \right ] \right ), \rho \right\}\bigg].
 \end{split}
\end{equation}

We thus see that $\mathcal{L}_{L}$ is another term of Lindblad form, in which the underlying rate matrix remains the same, but the basis of operators $\left\{A_n\right\}$ has been modified according to
\begin{equation}
A_{n} \to B_{n} = \phi \left [ A_{n} \right ].
\end{equation}
We can therefore write
\begin{equation}
\Gamma^{L;B} = \Gamma^{D;A}.
\end{equation}
Since this new term possesses a form of Lindblad type, with a physical rate matrix (under the assumption that the original rate matrix is physical), then there are no concerns about the physical nature of this new term. If we wish to express $\mathcal{L}_{L}$ as a Lindblad dissipator in the original basis, we can make use of the previously stated change of basis law to write
\begin{equation}
\Gamma^{L;A} = M \Gamma^{L;B} M^{\dagger} = M \Gamma^{D;A} M^{\dagger},
\end{equation}
or, for simplicity,
\begin{equation}
\Gamma^{L} = M \Gamma^{D} M^{\dagger},
\end{equation}
where $M$ is the matrix representation of $\phi$ in the original basis.

\subsection{The Form of the Matrix Element}

To evaluate $M$ explicitly, we recall 
\begin{equation}
\mathcal{L}_{H} = -i \left [ H, \cdot \right ] \equiv -i \text{ad}_{H},
\end{equation}
so that
\begin{equation}
\phi = \mathcal{T} \left \{ \exp \left ( -i\int_{0}^{t_{\text{op}}} \text{ad}_{H} \left ( t \right ) dt \right ) \right \},
\end{equation}
Since $M$ is the matrix representation of $\phi$, and since we have defined the object
\begin{equation}
\Omega_{mn} = \langle \langle A_{m} || \text{ad}_{H} || A_{n} \rangle \rangle = -i \sum_{p} H_{p} g_{pmn}
\end{equation}
to be the matrix representation of $\text{ad}_{H}$, we can write
\begin{equation}
M = \mathcal{T} \left \{ \exp \left ( -i \int_{0}^{t_{\text{op}}} \Omega \left ( t \right ) dt \right ) \right \},
\end{equation}
This follows simply from the fact that if one operator is defined as the exponential of another operator, consistency requires that their matrix representations obey the same relationship. 

However, one may be concerned whether this fact still holds in the presence of the time-ordering symbol. To remove any doubt regarding this, we can return to our original equation to be solved,
\begin{equation}
\phi e^{\mathcal{L}_{D}} = e^{\mathcal{L}_{L}} \phi.
\end{equation}
However, we now choose to approximate the total coherent evolution as a sequence of many small coherent evolutions, such that
\begin{equation}
\phi = \mathcal{T} \left \{ \exp \left ( \int_{0}^{t_{\text{op}}} \mathcal{L}_{H} \left ( t \right ) dt \right ) \right \}
\end{equation}
becomes
\begin{equation}
\phi  \to \prod_{n=0}^{N-1} \exp \left ( \tau \mathcal{L}_{H}^{n} \right ) ~ \equiv ~ \prod_{n=0}^{N-1} \phi^{(n)}
\end{equation}
where we have defined
\begin{equation}
\mathcal{L}_{H}^{n} = \mathcal{L}_{H} \left ( t_{n} \right ) = \mathcal{L}_{H} \left ( n \tau \right )
\end{equation}
for some small Trotter step $\tau$. We must therefore now solve
\begin{equation}
\left [ \prod_{n=0}^{N-1} \exp \left ( \tau \mathcal{L}_{H}^{n} \right ) \right ] e^{\mathcal{L}_{D}} = e^{\mathcal{L}_{L}} \left [ \prod_{n=0}^{N-1} \exp \left ( \tau \mathcal{L}_{H}^{n} \right ) \right ].
\end{equation}

To solve this new equation, we must commute our noise term to the left past each individual infinitesimal coherent term in the product. The first such commutation transforms the noise according to
\begin{equation}
\Gamma^{D} \to  M_{0}\Gamma^{D}M_{0}^{\dagger},
\end{equation}
where the factor $M_{0}$ correspond to commuting past the coherent evolution $\phi^{(0)}$. Likewise, after commuting past the second term, we have
\begin{equation}
\Gamma^{D} \to  M_{1} M_{0}\Gamma^{D}M_{0}^{\dagger}M_{1}^{\dagger}.
\end{equation}
Clearly, after commuting past all of the terms, we have
\begin{equation}
\Gamma^{D} \to \Gamma^{L} = \left [ \prod_{n=0}^{N-1} M_{n} \right ]  \Gamma^{D}\left [ \prod_{n=0}^{N-1} M_{n} \right ] ^{\dagger}.
\end{equation}
We thus identify the full $M$ transformation as
\begin{equation}
M = \prod_{n=0}^{N-1} M_{n}.
\end{equation}
We note that each $M_{n}$ is necessarily the matrix element of the infinitesimal operation $\phi^{(n)}$, by our previous result.

Now, since each $\phi^{(n)}$ is given in terms of a simple exponential, with no time-ordering symbol, there is no ambiguity in writing
\begin{equation}
\phi^{(n)} = e^{-i \tau \text{ad}_{H} \left ( t_{n} \right )} ~ \Rightarrow ~M_{n} =  e^{-i \tau \Omega \left (t_{n} \right )},
\end{equation}
thus allowing us to write
\begin{equation}
M = \prod_{n=0}^{N-1} e^{-i \tau \Omega \left (t_{n} \right )},
\end{equation}
If we now restore the continuum limit, our infinite product becomes the time-ordered exponential 
\begin{equation}
M = \mathcal{T} \left \{ \exp \left ( -i \int_{0}^{ t_{ \text{op} } } \Omega\left ( t \right ) dt    \right )   \right \},
\end{equation}
leading us to precisely the same result as before.

\subsection{Passing to the Right instead of the Left}

We have so far focused on solving the equation
\begin{equation}
\phi e^{\mathcal{L}_{D}} = e^{\mathcal{L}_{L}}\phi,
\end{equation}
which leads to the transformation of the rate matrix
\begin{equation}
\Gamma^{D} ~ \to ~ \Gamma^{L} ~ \equiv ~ M\Gamma^{D}M^\dagger
\end{equation}
This corresponds to commuting a noise term to the left of a coherent term. We may naturally ask what happens when we instead commute a noise term to the right of a coherent term, so that we wish to solve the equation
\begin{equation}
\phi e^{\mathcal{L}_{R}} = e^{\mathcal{L}_{D}} \phi,
\end{equation}

Fortunately, no extra work is required in this case. We note that mathematically, we can solve the above equation precisely as before, except with
\begin{equation}
\mathcal{L}_{D} \to \mathcal{L}_{R} ~;~ \mathcal{L}_{L} \to \mathcal{L}_{D}
\end{equation}
This then leads to
\begin{equation}
\Gamma^{D} ~ = ~ M\Gamma^{R}M^\dagger,
\end{equation}
with M defined in exactly the same way. This expression can then be rearranged to find
\begin{equation}
\Gamma^{R} ~ = ~ M^\dagger\Gamma^{D}M.
\end{equation}

\subsection{A Corollary Involving Commutators}

We briefly mention here a corollary of the results derived above. In particular, these results can be used to argue that the commutator of a coherent super-operator with a Lindblad dissipation super-operator yields another Lindblad dissipation super-operator.

To show this, we consider a time-independent, coherent quantum operation
\begin{equation}
\phi = e^{ \tau \mathcal{L}_{H}}.
\end{equation}
If we commute a noise term to the left of this quantum operation, we know that it will transform according to
\begin{equation}
\mathcal{L}_{L} = \phi \mathcal{L}_{D} \phi^{-1} = e^{ \tau \mathcal{L}_{H}} \mathcal{L}_{D} e^{- \tau \mathcal{L}_{H}}.
\end{equation}
This can be rewritten as
\begin{equation}
\mathcal{L}_{L} = e^{ \tau \text{ad}_{\mathcal{L}_{H}}} \mathcal{L}_{D} = \mathcal{L}_{D} + \tau \left [ \mathcal{L}_{H}, \mathcal{L}_{D} \right ] + \ldots 
\end{equation}
Similarly, we know that the rate matrix of $\mathcal{L}_{L}$ should be given by
\begin{equation}
\Gamma^{L} = M \Gamma^{D} M^{\dagger} = e^{ - i \tau \Omega} \Gamma^{D} e^{ + i \tau \Omega}.
\end{equation}
This can be likewise rewritten as
\begin{equation}
\Gamma^{L} =  e^{ - i \tau \text{ad}_{\Omega}} \Gamma^{D} = \Gamma^{D} + \tau \left [ -i \Omega, \Gamma^{D} \right ] + \ldots
\end{equation}

Now, for a given choice of basis, Lindblad dissipator terms are in one to one correspondence with their rate matrices. The rate matrix $\Gamma^{L}$ corresponds to the dissipator $\mathcal{L}_{L}$. Furthermore, adding two Lindblad dissipators results in another dissipator, whose rate matrix is the sum of the rate matrices of the original dissipators. For the association between $\Gamma^{L}$ and $\mathcal{L}_{L}$ to hold for all $\tau$, each of the terms in the expansion of $\mathcal{L}_{L}$ must correspond to a dissipator of Lindblad form, whose rate matrix is given by the associated term in the expansion of $\Gamma^{L}$. For this reason, if we commute a coherent super-operator with a Lindblad dissipator,
\begin{equation}
\mathcal{L}_{T} = \left [ \mathcal{L}_{H}, \mathcal{L}_{D} \right ],
\end{equation}
we see that the resulting object $\mathcal{L}_{T}$ takes the form of a Lindblad dissipator, whose rate matrix is given by
\begin{equation}
\Gamma^{T} =  -i \left [\Omega, \Gamma^{D} \right ].
\end{equation}

We emphasize, however, that while $\mathcal{L}_{T}$ has the structural form of a Lindblad dissipator, the rate matrix $\Gamma^{T}$ will not, in general, be positive semi-definite (it will, however, always be Hermitian, so long as $\Gamma^{D}$ is Hermitian). It is only when we sum up the full expansion that we find $\Gamma^{L}$ to be positive semi-definite.

%% file: Sections/appSepDeriv.tex
\section{Derivation of The Separated Noise}
\label{sec:appSepDeriv}

Here we provide a derivation of the general form of the separated noise. We start by deriving the closed-form integral expression for the separated noise, and then proceed to show that it satisfies the desired differential equation. We also provide an alternate derivation of the separated noise which is valid in the time-independent case.

\subsection{Derivation of the Integral Expression}

To derive the integral expression defining the separated noise, we give two alternate derivations. The first derivation involves Trotterizing the quantum operation, while the second derivation invokes the interaction picture.

\subsubsection{Derivation through Trotterization}

To proceed with the first derivation, we consider a noisy quantum operation
\begin{equation}
\Phi = \mathcal{T} \left \{ \exp \left ( \int_{0}^{t_{\text{op}}} \mathcal{L} \left ( t \right ) dt \right ) \right \},
\end{equation}
with
\begin{equation}
\begin{split}
& \mathcal{L} \\ \equiv~ & \mathcal{L}_{H} + \mathcal{L}_{D} \\
\equiv~ & -i \left [ H, \cdot \right ] + \sum_{n,m} \Gamma^{D}_{nm}\left[A_n \left ( \cdot \right ) A_m^\dagger-\frac{1}{2}\left\{A_m^\dagger A_n, \left ( \cdot \right ) \right\}\right]
\end{split}
\end{equation}
We now decompose this operation into a sequence of $N$ ``Trotter'' steps of size $\tau = t_{\text{op}}/N$ (we note that this decomposition is merely a formal exercise, since while the implementation of the coherent dynamics is controlled by the operator of the quantum hardware, noise is introduced through the environment, so in practice one would not really be able to physically Trotterize the noise terms). After performing such a decomposition, we have
\begin{equation}
\Phi = \prod_{n=0}^{N-1} \mathcal{T} \left \{ \exp \left ( \int_{t_{n}}^{t_{n+1}} \mathcal{L} \left ( t \right ) dt \right ) \right \},
\end{equation}
where we have defined
\begin{equation}
t_{n} = n \tau.
\end{equation}
Since we take the Trotter parameter $\tau$ to be appropriately small, we can approximate
\begin{equation}
\int_{t_{n}}^{t_{n+1}} \mathcal{L} \left ( t \right ) dt ~\approx~ \tau \mathcal{L} \left ( t_{n} \right ) 
\end{equation}
in order to find
\begin{align}
\Phi_{\text{op}} &\approx \prod_{n=0}^{N-1}   \exp \left ( \tau \mathcal{L} \left ( t_{n} \right )\right ) \nonumber \\
&= \prod_{n=0}^{N-1}   \exp \left ( \tau \mathcal{L}_{H} \left ( t_{n} \right ) + \tau\mathcal{L}_{D} \left ( t_{n} \right ) \right ).
\end{align}
We now split apart the exponential, to find
\begin{equation}
\Phi ~\approx~ \prod_{n=0}^{N-1} \exp \left ( \tau \mathcal{L}_{H}^{n} \right )  \exp \left ( \tau \mathcal{L}_{D}^{n}  \right ),
\end{equation}
where we have defined
\begin{equation}
\mathcal{L}^{n}_{H} \equiv \mathcal{L}_{H} \left ( t_{n} \right ) ~;~ \mathcal{L}^{n}_{D} \equiv \mathcal{L}_{D} \left ( t_{n} \right )
\end{equation}
We note that we have placed the noise term on the right.

At this point, we may wish to commute all of the noise terms to the left, resulting in an expression which is equivalent to a coherent evolution, followed by a noise term. However, because the coherent terms increase in time from right to left, we will actually find it easier to begin by pushing all of the noise terms to the right, only to commute the entire noise term all the way to the left as the final step. The first noise term does not need to be commuted past any coherent terms, since it is already to the right of every other term in the expression. The next term must be commuted past one single coherent term, the following noise term after that must be commuted past two coherent terms, and so on. As a result, we find
\begin{equation}
\Phi ~\approx~ \prod_{n=0}^{N-1} \exp \left ( \tau \mathcal{L}_{H}^{n} \right ) \prod_{n=0}^{N-1} \exp \left ( \tau \mathcal{L}_{R}^{(n)} \right ),
\end{equation}
where $\mathcal{L}_{R}^{(n)}$ is the noise term which results from commuting $\mathcal{L}_{D}^{n}$ to the right past $n$ coherent terms. In particular, $\mathcal{L}_{R}^{(n)}$ results from commuting $\mathcal{L}_{D}^{n}$ past the object
\begin{equation}
\phi^{(n)} \equiv \prod_{m=0}^{n-1} \exp \left ( \tau \mathcal{L}_{H}^{m} \right ).
\end{equation}

After commuting all of the noise terms to the right, as a next step we would like to combine all of these modified noise terms into a single exponential
\begin{equation}
\prod_{n=0}^{N-1} \exp \left ( \tau \mathcal{L}_{R}^{(n)} \right ) \to  \exp \left ( \tau  \sum_{n=0}^{N-1} \mathcal{L}_{R}^{(n)} \right )
\end{equation}
Since the sum of the objects $\mathcal{L}_{R}^{(n)}$ yields another Lindblad term, we could then make the association
\begin{equation}
t_{\text{op}} \mathcal{L}_{F} = \sum_{n=0}^{N-1}  \tau \mathcal{L}_{R}^{(n)}
\end{equation}
for some $\mathcal{L}_{F}$ which is of pure Lindblad type (this Lindblad term would correspond to the ``forward'' noise, which would correspond to the separated noise if it were modeled as occurring before, rather than after, the coherent evolution). However, here we must be careful about the magnitude of the individual noise terms. To second order in the BCH expansion, the various exponential terms can be recombined according to
\begin{equation}
t_{\text{op}}\mathcal{L}_{F} ~\approx~ \tau \sum_{n} \mathcal{L}_{R}^{(n)} + \frac{1}{2} \tau^{2} \sum_{n < m} \left [ \mathcal{L}_{R}^{(n)}, \mathcal{L}_{R}^{(m)} \right ].
\end{equation}
While it is true that
\begin{equation}
\tau^{2} ~\sim~ 1/N^{2},
\end{equation}
it is also true that the number of commutator terms is quadratic in the number of Trotter steps. Thus, we cannot immediately guarantee the smallness of the second term (or any higher-order terms, for that matter). This is potentially problematic, since it can be shown that the commutator of two Lindblad super-operators does not yield another Lindblad super-operator, but rather results in a term of mixed type, with both Lindblad and Hamiltonian contributions. Thus, we see that we must make the assumption that the noise terms themselves are appropriately small, so that we can disregard the commutator term due to being quadratically small in the noise strength.

We see then that to lowest order in the noise strength, it is possible to recombine the arguments of the exponentials in the product, in order to indeed find
\begin{equation}
t_{\text{op}} \mathcal{L}_{F} ~\approx~ \sum_{n=0}^{N-1}  \tau \mathcal{L}_{R}^{(n)}.
\end{equation}
Our expression for the quantum map can now be written
\begin{equation}
\Phi ~\approx~ \left [ \prod_{n=0}^{N-1} \exp \left ( \tau \mathcal{L}_{H}^{n} \right ) \right ] \exp \left ( t_{\text{op}} \mathcal{L}_{F} \right ).
\end{equation}
We must now examine what happens in the limit as we take the size of the Trotter step to zero. The coherent term on the left approaches a time-ordered exponential
\begin{equation}
\prod_{n=0}^{N-1} \exp \left ( \tau \mathcal{L}_{H}^{n} \right ) \to \mathcal{T} \left \{ \exp \left ( \int_{0}^{t_{\text{op}}} \mathcal{L}_{H} \left ( t \right ) dt \right ) \right \},
\end{equation}
which is just the coherent evolution in the absence of any noise. For the forward noise, we have
\begin{equation}
t_{\text{op}}  \mathcal{L}_{F} = \sum_{n=0}^{N-1}  \tau \mathcal{L}_{R}^{(n)} \to \int_{0}^{t_{\text{op}}} \mathcal{L}_{R} \left ( s \right ) ds,
\end{equation}
where the term $\mathcal{L}_{R}\left ( s \right )$ results from commuting the bare noise past the coherent evolution up until time $s$, which in the continuum limit goes to
\begin{equation}
\begin{split}
\phi^{(n)} = & \prod_{m=0}^{n-1} \exp \left ( \tau \mathcal{L}_{H}^{m} \right ) \to \\ \phi\left ( s \right ) = & \mathcal{T} \left \{ \exp \left ( \int_{0}^{s} \mathcal{L}_{H} \left ( t \right ) dt \right ) \right \}.
\end{split}
\end{equation}

Using our results for the effects of commuting a noise term past a coherent quantum operation to the right, we have
\begin{equation}
\mathcal{L}_{R}\left ( s \right ) = \phi^{-1} \left ( s \right ) \mathcal{L}_{D}\left ( s \right ) \phi\left ( s \right ),
\end{equation}
which leads to the expression
\begin{equation}
t_{\text{op}}  \mathcal{L}_{F} = \int_{0}^{t_{\text{op}}} \phi^{-1} \left ( s \right ) \mathcal{L}_{D}\left ( s \right ) \phi\left ( s \right ),
\end{equation}
Finally, commuting the forward noise term all the way to the left of the full coherent evolution in order to get the separated noise, we find
\begin{equation}
t_{\text{op}}  \mathcal{L}_{S} = \phi  \left [ \int_{0}^{t_{\text{op}}} \phi^{-1} \left ( s \right ) \mathcal{L}_{D} \left ( s \right ) \phi\left ( s \right ) \right ] \phi^{-1}.
\end{equation}
Since we know the effect that commuting a noise term past a coherent gate has on its underlying rate matrix, this allows us to write our result as
\begin{equation}
t_{\text{op}}  \Gamma^{S} =  M \left [ \int_{0}^{t_{\text{op}}} M^{\dagger} \left ( s \right ) \Gamma^{D} \left ( s \right )  M\left ( s \right ) ds \right ] M^{\dagger},
\end{equation}
where, once again,
\begin{equation}
M \left ( s \right ) = \mathcal{T} \left \{ \exp \left ( -i \int_{0}^{s} \Omega \left ( t \right ) dt \right )   \right \},
\end{equation}
which is the previously stated result.

\subsubsection{Derivation with the Interaction Picture}

Here we give an alternate derivation of the integral expression, through the use of the interaction picture. We begin with the time-dependent generator
\begin{equation}
\mathcal{L} = \mathcal{L}_{H} + \mathcal{L}_{D}.
\end{equation}
We identify this as the evolution operator for the Schr\"odinger picture,
\begin{equation}
\frac{d}{dt}\rho_{S} = \mathcal{L} \left [  \rho_{S} \right ].
\end{equation}
We then define the interaction picture density matrix by
\begin{equation}
\rho_{I} \left ( s \right ) =  \phi^{-1} \left ( s \right ) \left [ \rho_{S} \left ( s \right ) \right ]   \Leftrightarrow \rho_{S} \left ( s \right ) = \phi \left ( s \right ) \left [ \rho_{I} \left ( s \right ) \right ],
\end{equation}
where $\phi \left ( s \right )$ is again just the coherent part
\begin{equation}
\phi \left ( s \right ) = \mathcal{T} \left \{ \exp \left ( \int_{0}^{s} \mathcal{L}_{H} \left ( t \right ) dt \right ) \right \}.
\end{equation}
We also define the noise term time-evolved in the interaction picture as
\begin{equation}
\mathcal{L}^{I}_{D} \left ( s \right ) = \phi^{-1}\left ( s \right ) \mathcal{L}_{D} \left ( s \right ) \phi \left ( s \right ).
\end{equation}
If we make use of the fact that
\begin{equation}
\frac{d}{ds} \phi \left ( s \right ) = \mathcal{L}_{H} \left ( s \right ) \phi \left ( s \right )
\end{equation}
which follows from the definition of $\phi \left ( s \right )$ in terms of the time-ordered exponential of $\mathcal{L}_{H}$, as well as the fact that $\mathcal{L}_{H}$ is anti-Hermitian, then using all of these definitions, it is straightforward to verify that
\begin{equation}
\frac{d}{dt}\rho_{I} = \mathcal{L}^{I}_{D} \left [ \rho_{I} \right ].
\end{equation}
Solving this equation formally, we get
\begin{equation}
\rho_{I} \left ( t_{\text{op}} \right ) = \mathcal{T} \left \{ \exp \left ( \int_{0}^{t_{\text{op}}} \mathcal{L}^{I}_{D}  \left ( s \right ) ds \right ) \right \} \rho_{I} \left ( 0 \right )
\end{equation}

Now, from our previous results regarding the effects of commuting a noise term past a coherent term, we know that the object
\begin{equation}
\mathcal{L}^{I}_{D} \left ( s \right ) = \phi^{-1}\left ( s \right ) \mathcal{L}_{D} \left ( s \right ) \phi \left ( s \right )
\end{equation}
is yet another noise term, with the same overall noise strength as $\mathcal{L}_{D}$. For this reason, we can assume that $\mathcal{L}^{I}_{D}$ is appropriately small, in the same sense that the original $\mathcal{L}_{D}$ is small. In particular, we will assume that it is small enough that we can effectively take the commutator of $\mathcal{L}^{I}_{D}$ with itself at different times to be zero. In this way, our approximation that the noise is appropriately small leads to an approximation in which we can simply ignore the time-ordering symbol. This then leads us to the expression
\begin{equation}
\rho_{I} \left ( t_{\text{op}} \right ) = \exp \left ( \int_{0}^{t_{\text{op}}} \mathcal{L}^{I}_{D}  \left ( s \right ) ds \right ) \rho_{I} \left ( 0 \right ).
\end{equation}

We can now transform back to the Schr\"odinger picture, to get
\begin{equation}
\rho_{S} \left ( t_{\text{op}} \right ) = \phi  \exp \left ( \int_{0}^{t_{\text{op}}} \mathcal{L}^{I}_{D}  \left ( s \right ) ds \right ) \rho_{S} \left ( 0 \right ),
\end{equation}
since the two pictures coincide at time zero. If we define
\begin{equation}
t_{\text{op}} \mathcal{L}_{F} = \int_{0}^{t_{\text{op}}} \mathcal{L}^{I}_{D}  \left ( s \right ) ds,
\end{equation}
then we identify
\begin{equation}
\Phi = \phi e^{t_{\text{op}} \mathcal{L}_{F}}
\end{equation}
as the full quantum map in the Schrodinger picture. Finally, commuting the noise term to the left past the coherent term, we have
\begin{equation}
t_{\text{op}} \mathcal{L}_{F} \to t_{\text{op}} \mathcal{L}_{S} = \phi \mathcal{L}_{F} \phi^{-1} =  \phi \left [ \int_{0}^{t_{\text{op}}} \mathcal{L}^{I}_{D}  \left ( s \right ) ds \right ] \phi^{-1}.
\end{equation}
Using the explicit form of $\mathcal{L}^{I}_{D}$, we find
\begin{equation}
\mathcal{L}_{S} = \frac{1}{t_{\text{op}}} \phi \left [ \int_{0}^{t_{\text{op}}} \phi^{-1}  \left ( s \right ) \mathcal{L}_{D} \left ( s \right ) \phi  \left ( s \right )  ds \right ] \phi^{-1}
\end{equation}
which is once again the previously stated result.

\subsection{Derivation of the Differential Equation}

The integral expression we have derived for the separated noise may not always be practical to compute. We therefore would like a differential equation which can be used to actually solve for the separated noise. To this end, we define, as before
\begin{align}
Q \left ( s \right ) &= s \Gamma^{S} \left ( s \right ) \nonumber \\
&=  M \left ( s \right ) \left [ \int_{0}^{s} M^{\dagger} \left ( r \right ) \Gamma^{D}\left ( r \right )  M\left ( r \right ) dr \right ] M^{\dagger}\left ( s \right ).
\end{align}
It is clear from this definition that
\begin{equation}
Q \left ( s = 0 \right ) = 0.
\end{equation}
If we can find a differential equation for $Q$, then this equation can be integrated from zero up until $t_{\text{op}}$, in order to find the rate matrix of the separated noise.

To begin, we define
\begin{equation}
I \left ( s \right ) \equiv \int_{0}^{s} M^{\dagger} \left ( r \right ) \Gamma^{D} \left ( r \right )  M\left ( r \right ) dr,
\end{equation}
which allows us to write
\begin{equation}
Q \left ( s \right ) = M \left ( s \right ) I\left ( s \right )  M^{\dagger}\left ( s \right ).
\end{equation}
Differentiating $Q$, we find
\begin{equation}
\frac{dQ} {ds} = \left [ \frac{dM} {ds} \right ] I M^{\dagger} + M \left [ \frac{dI} {ds} \right ] M^{\dagger} + MI \left [ \frac{dM} {ds} \right ]^{\dagger}.
\end{equation}
Notice that since $M$ is a finite-dimensional matrix, there is no potential ambiguity about whether the actions of differentiation and conjugation commute. Clearly, we have
\begin{equation}
\frac{dI} {ds} = M^{\dagger} \left ( s \right ) \Gamma^{D} \left ( s \right )  M\left ( s \right ),
\end{equation}
which is simply a result of the fundamental theorem of calculus. Furthermore, due to its definition in terms of a time-ordered exponential, we have
\begin{equation}
\frac{dM} {ds} = -i \Omega \left ( s \right ) M \left ( s \right ).
\end{equation}

Combining these various pieces together, we have
\begin{equation}
\frac{dQ} {ds} = -i \Omega  M I M^{\dagger} + M M^{\dagger} \Gamma^{D}  M M^{\dagger} + i M I M^{\dagger} \Omega,
\end{equation}
where all of the time-dependent objects are evaluated at the same time $s$, and where we have made use of the fact that $\Omega$ is hermitian. Since $M$ is unitary, this can be simplified to
\begin{equation}
\frac{dQ} {ds} = -i \Omega  M I M^{\dagger} + \Gamma^{D}  + i M I M^{\dagger} \Omega.
\end{equation}
Using the original definition of $Q$, this further reduces to
\begin{equation}
\frac{dQ} {ds} = -i \Omega  Q + \Gamma^{D}  + i Q \Omega,
\end{equation}
or, alternatively
\begin{equation}
\frac{dQ} {ds} = -i \left [ \Omega,  Q \right ] + \Gamma^{D} = \xi \left [ Q \right ] + \Gamma^{D},
\end{equation}
which is the previously stated result.

\subsection{Alternate Derivation of the Separated Noise in the Time-Independent Case}

Here we provide an alternative derivation of the formula for the separated noise in the time-independent case. This will provide an independent check of the accuracy of our results, as well as provide an alternative expression for computing the separated noise.

To begin, we again write the quantum operation in question as
\begin{equation}
\Phi = \mathcal{T} \left \{ \exp \left ( \int_{0}^{t_{\text{op}}} \mathcal{L} \left ( t \right ) dt \right ) \right \}.
\end{equation}
Since the Lindblad generator is no longer time-dependent in this case, we can write simply
\begin{equation}
\Phi =  \exp \left ( t_{\text{op}} \mathcal{L}  \right ) = \exp \left ( t_{\text{op}} \mathcal{L}_{H} + t_{\text{op}} \mathcal{L}_{D} \right ).
\end{equation}
Our hope now is to be able to find an object $\mathcal{L}_{S}$ such that, to first order in the noise strength, we have
\begin{equation}
\exp \left ( t_{\text{op}} \mathcal{L}_{H} + t_{\text{op}} \mathcal{L}_{D} \right ) \approx \exp \left ( t_{\text{op}} \mathcal{L}_{S} \right ) \exp \left ( t_{\text{op}} \mathcal{L}_{H} \right ).
\end{equation}
It is clear then that to find an expression for the separated noise in this time-independent case, we must find a solution to the equation
\begin{equation}
e^{X+Y} = e^W e^X
\end{equation}
for $W$ in terms of $X$ and $Y$, under the assumption that $Y$ is sufficiently small in comparison to $X$. 

A recursive formula \cite{kimura} for the exponentiation of $W$ can be written 
\begin{equation}
e^W = \sum_{m=0}^{\infty} \frac{1}{m!} V_{m},
\end{equation}
where
\begin{equation}
V_{0} = I ~;~ V_{1} = Y ~;~ V_{m+1} = \left [ X, V_{m} \right ] + Y V_{m}
\end{equation}
When only retaining terms to lowest order in Y, this recursion relation simplifies to
\begin{equation}
V_{0} = I ~;~ V_{m \neq 0} = \Lambda^{(m-1)} \left ( Y \right ),
\end{equation}
where
\begin{equation}
\Lambda\left ( Y \right ) \equiv \left [ X, Y \right ]
\end{equation}
such that $\Lambda^{(m-1)}$ is the $(m-1)$ times nested commutator under $X$. 
We can thus write
\begin{equation}
e^W = I + \sum_{m=1}^{\infty} \frac{1}{m!} \Lambda^{(m-1)} \left ( Y \right ),
\end{equation}
or, alternatively,
\begin{equation}
e^W = I + \sum_{m=0}^{\infty} \frac{1}{(m+1)!} \Lambda^{m} \left ( Y \right ).
\end{equation}

Now, assuming that $W$ has some expansion in powers of the overall scale of $Y$, we can write
\begin{equation}
W = w_{0} + \lambda w_{1} + \lambda^{2} w_{2} + ...
\end{equation}
where $\lambda$ is the overall scale of $Y$. From the original defining equation, it is clear that when $Y$ vanishes, such that $\lambda = 0$, we must have
\begin{equation}
\begin{split}
e^{X} = e^W e^X ~\Rightarrow ~ e^W = I = e^0 ~\Rightarrow~ \\
W \left ( \lambda = 0 \right ) = 0 ~\Rightarrow~ w_{0} = 0.
\end{split}
\end{equation}
Thus, $W$ itself must already be of order $\lambda$. We therefore find that
\begin{equation}
\begin{split}
e^W \approx I + & W + \frac{1}{2} W^2 + ... \\
\approx I + \left ( \lambda w_{1} + \lambda^{2} w_{2} + ... \right ) & + \frac{1}{2} \! \left ( \lambda w_{1} \! + \lambda^{2} w_{2} + \! ... \right )^2 + ... \\
\approx I & + \lambda w_{1} + ...
\end{split}
\end{equation}

Comparing the two expressions we have found for the exponentiation of $W$ to first order, we see that we can make the association
\begin{equation}
\lambda w_{1} ~ \to ~ \sum_{m=0}^{\infty} \frac{1}{(m+1)!} \Lambda^{m} \left ( Y \right ),
\end{equation}
which, to lowest order, implies
\begin{equation}
W \approx \sum_{m=0}^{\infty} \frac{1}{(m+1)!} \Lambda^{m} \left ( Y \right ).
\end{equation}
We note that since $\Lambda$ is a bounded linear operator, and since the coefficients in this series decay at least as quickly as the series expansion of the exponential function, then there should be no questions regarding the convergence of this series.

We thus see that if we define the adjoint action
\begin{equation}
\Lambda\left ( Y \right ) \equiv \left [ X, Y \right ]
\end{equation}
then $W$ is given according to
\begin{equation}
W = \mathcal{F} \left [ Y \right] 
\end{equation}
where $\mathcal{F}$ is defined by the series expansion
\begin{equation}
\mathcal{F} \left [ Y \right] ~=~ \sum_{m=0}^{\infty} \frac{1}{(m+1)!} \Lambda^{m} \left ( Y \right ).
\end{equation}
Applying this result to the problem at hand, we find
\begin{equation}
\mathcal{L}_{S} = \mathcal{F} \left [ \mathcal{L}_{D} \right],
\end{equation}
where we now have
\begin{equation}
\Lambda \left ( \mathcal{L}_{D} \right ) = t_{\text{op}} \left [ \mathcal{L}_{H}, \mathcal{L}_{D} \right ] 
\end{equation}

Having found a formal expression for $\mathcal{L}_{S}$ to first order in the noise strength, we must evaluate the commutator between $\mathcal{L}_{H}$ and $\mathcal{L}_{D}$. We know from Appendix~\ref{sec:appCommuteGate} that this commutator produces another super-operator of Lindblad form,
\begin{equation}
t_{\text{op}}\left [ \mathcal{L}_{H}, \mathcal{L}_{D} \right ] = t_{\text{op}}\mathcal{L}_{T},
\end{equation}
with a rate matrix given according to
\begin{equation}
\Gamma^{T} = -i \left [ \Omega, \Gamma^{D} \right ] = \xi \left [ \Gamma^{D}\right ]
\end{equation}
with $\xi$ and $\Omega$ defined as before. We recall that the rate matrix $\Gamma^{T}$ is always Hermitian, however it is not positive semi-definite (it thus only superficially takes the form of a Lindblad dissipator).

We thus see that the action $\Lambda \left ( \mathcal{L}_{D}\right )$ corresponds to an object which superficially takes the form of a Lindblad noise term, with a rate matrix transformed under the map $\xi$. Since acting $\Lambda$ on a Lindblad noise term produces another Lindblad noise term, the same must hold true for any higher-order power of $\Lambda$. Therefore, the full expression for the transformed noise,
\begin{equation}
\mathcal{L}_{S} = \mathcal{F}\left [ \mathcal{L}_{D} \right] =  \sum_{m=0}^{\infty} \frac{1}{(m+1)!} \Lambda^{m} \left (\mathcal{L}_{D} \right )
\end{equation}
must likewise correspond to a term of Lindblad form, since the sum of two Lindblad terms is another Lindblad term. It is clear that the transformation of the underlying rate matrix under a higher-order power of $\Lambda$ is found via the corresponding higher-order power of the map $\xi$,
\begin{equation}
\Lambda^{m} \left ( \mathcal{L}_{D} \right ) \to t_{\text{op}}^{m} \xi^{m} \left [ \Gamma^{D}\right ],
\end{equation}
and so we can alternatively write this transformation as
\begin{equation}
\Gamma^{S} = \mathcal{K} \left [ \Gamma^{D} \right] = \sum_{m=0}^{\infty} \frac{ t_{\text{op}}^{m}}{(m+1)!} \xi^{m} \left [ \Gamma^{D} \right ],
\end{equation}
which is the previously stated result.

To show that this alternate expression is indeed equivalent to the original one in the time-independent case, we consider again the object
\begin{equation}
Q \left ( s \right ) = s \Gamma^{S} \left ( s \right ) = \sum_{m=0}^{\infty} \frac{s^{(m+1)}}{(m+1)!} \xi^{m} \left ( \Gamma^{D} \right ).
\end{equation}
Differentiating this expression, we have
\begin{equation}
\begin{split}
\frac{dQ}{d s}  & = \frac{d}{d s} \sum_{m=0}^{\infty} \frac{ s^{(m+1)}}{(m+1)!} \xi^{m} \left ( \Gamma^{D} \right ) \\ & =  \sum_{m=0}^{\infty} \frac{ s^{m}}{m!} \xi^{m} \left ( \Gamma^{D} \right ).
\end{split}
\end{equation}
The second summation is simply the exponentiation of the quantity $s \xi$, and thus 
\begin{equation}
\frac{d} {d s} Q  = e^{ s \xi } \left [ \Gamma^{D} \right ] = e^{-i  \Omega s } \Gamma^{D} e^{+i \Omega s},
\end{equation}
which follows from the fact that $\xi$ is the infinitesimal unitary transformation
\begin{equation}
\xi \left [ \Gamma^{D} \right ] ~\equiv ~ -i \left [ \Omega, \Gamma^{D} \right ].
\end{equation}
Integrating the resulting derivative expression from zero up until $t_{\text{op}}$, we find
\begin{equation}
Q \left ( t_{\text{op}} \right ) ~=~  t_{\text{op}} \Gamma^{S}  ~=~  \int_{0}^{t_{\text{op}}} e^{-i  \Omega s } \Gamma^{D} e^{+i \Omega s} ds,
\end{equation}
which is indeed equivalent to the original expression, in the time-independent case.

%% file: Sections/appPSD.tex
\section{Physical Nature of the Separated Noise}
\label{sec:appPhys}

Here we briefly discuss the conditions for the separated noise to be physical, as well as the relationship between the total noise strength of the separated noise and the total noise strength of the original noise.

\subsection{Analytic Argument}

Our expression for the separated noise is given according to
\begin{equation}
\Gamma^{S} = \frac{1}{t_{\text{op}}} M \left [ \int_{0}^{t_{\text{op}}} M^{\dagger} \left ( s \right ) \Gamma^{D} \left ( s \right )  M \left ( s \right ) ds \right ] M^{\dagger}.
\end{equation}
For the separated noise to be physical, the rate matrix $\Gamma^{S}$ must be Hermitian and positive semi-definite. We emphasize that this requirement is strict, because the separated noise dynamics correspond to time-independent Lindblad noise -- in this case, it is known that the rate matrix must be Hermitian and PSD in order to represent physical dynamics.

It is relatively straight-forward to see why this should be the case, due to the general structure of the above transformation, so long as $\Gamma^{D}$ itself is Hermitian and PSD at all times (we will comment shortly on the more general case that $\Gamma^{D}$ is not necessarily PSD at all times). The integrand of the transformation corresponds to a unitary transformation of $\Gamma^{D}$, which does not alter the spectrum of $\Gamma^{D}$, nor does it alter whether $\Gamma^{D}$ is Hermitian. Therefore, assuming that the underlying hardware noise possesses these properties at all times, so will the integrand. Furthermore, adding together Hermitian, positive semi-definite matrices results in another Hermitian, positive semi-definite matrix, so the integral itself results in a rate matrix which is Hermitian and PSD. Lastly, performing the outer unitary transformation, and dividing by the positive parameter $t_{\text{op}}$, we have a rate matrix which is still PSD and Hermitian, and thus physical.

We can make the above argument somewhat more rigorous. Again, we note that because the matrix $M$ is unitary, the spectrum of the forward noise
\begin{equation}
\Gamma^{F} = \frac{1}{t_{\text{op}}} \int_{0}^{t_{\text{op}}} M^{\dagger} \left ( s \right ) \Gamma^{D}\left ( s \right )  M \left ( s \right )  ds
\end{equation}
will be the same as the spectrum of the separated noise. It is therefore sufficient to prove that the rate matrix of the forward noise is physical, which is to say that
\begin{equation}
\langle v | \Gamma^{F} | v \rangle \geq 0
\end{equation}
for any arbitrary vector $v$. Since we are working with a finite-dimensional Hilbert space, there is no ambiguity associated with pulling the inner product inside of the integral which defines $\Gamma^{F}$, and thus we find
\begin{equation}
\langle v | \Gamma^{F} | v \rangle =  \frac{1}{t_{\text{op}}}  \left [ \int_{0}^{t_{\text{op}}} \langle v | M^{\dagger} \left ( s \right ) \Gamma^{D}\left ( s \right )  M \left ( s \right ) | v \rangle ds \right ] . 
\end{equation}
If we now define
\begin{equation}
| w \left ( s \right ) \rangle =  M \left ( s \right ) | v \rangle ,
\end{equation}
then we have
\begin{equation}
\langle v | \Gamma^{F} | v \rangle =  \frac{1}{t_{\text{op}}}  \left [ \int_{0}^{t_{\text{op}}} \langle w \left ( s \right ) | \Gamma^{D}\left ( s \right )  | w \left ( s \right ) \rangle ds \right ] . 
\end{equation}
If we assume that the matrix $\Gamma^{D}$ is PSD at all times, we have
\begin{equation}
\langle w \left ( s \right ) | \Gamma^{D}\left ( s \right )  | w \left ( s \right ) \rangle \geq 0
\end{equation}
for all values of $s$. The integral above therefore corresponds to the integral over a strictly non-negative quantity, which of course must be non-negative. We thus indeed find
\begin{equation}
\langle v | \Gamma^{F} | v \rangle \geq 0.
\end{equation}

Because the spectrum of the forward noise and separated noise are the same, we can compute the total noise strength of the separated noise as
\begin{equation}
\overline{\gamma}^{S} = \text{Tr} \left [  \Gamma^{F} \right ]  = \frac{1}{t_{\text{op}}} \text{Tr} \left [\int_{0}^{t_{\text{op}}} M^{\dagger} \left ( s \right ) \Gamma^{D} \left ( s \right ) M \left ( s \right ) ds \right ]
\end{equation}
Again, since we are working with a finite-dimensional Hilbert space, we can pull the trace inside of the integral, to find
\begin{equation}
\overline{\gamma}^{S}  = \frac{1}{t_{\text{op}}} \int_{0}^{t_{\text{op}}} \text{Tr} \left [ M^{\dagger} \left ( s \right ) \Gamma^{D} \left ( s \right )  M \left ( s \right ) \right ] ds .
\end{equation}
The cyclic nature of the trace eliminates the unitary transformation, and we have
\begin{equation}
\overline{\gamma}^{S}  = \frac{1}{t_{\text{op}}} \int_{0}^{t_{\text{op}}} \overline{\gamma}^{D} \left ( s \right ) ds .
\end{equation}
The expression above tells us that the total noise strength of the separated noise is given by the time-average of the total noise strength of the underlying hardware noise. In the special case that the hardware noise is constant, the two noise strengths are precisely equal.

In general, whether or not the matrix $\Gamma^{F}$ is PSD will depend on the interplay between the time-dependent objects $M \left ( s \right )$ and $\Gamma^{D} \left ( s \right )$, and thus the interplay between the coherent and noisy dynamics. We do not attempt an analysis of this question in the general case. We note, however, that certain simple cirteria may be able to determine whether the matrix is PSD in some cases. For example, if a matrix is PSD, its trace must be non-negative. Thus, if the average strength of the original hardware noise is negative, the separated noise will necessarily be unphysical. 

We note an interesting corollary of the above considerations. We recall that the separated noise represents a method for approximating the original full dynamics,
\begin{equation}
\Phi \approx \exp \left ( t_{\text{op}}\mathcal{L}_{S} \right ) \phi.
\end{equation}
Since the unitary dynamics will always be physical, the physical nature of $\Phi$ depends entirely on the physical nature of the separated noise term. Therefore, in the weak noise limit, the physical nature of $\Phi$ will be determined by the same interplay between the coherent and noisy dynamics that determines whether the matrix $\Gamma^{F}$ is PSD.

\subsection{Further Numerical Analysis of the Physical Nature of the Separated Noise}

We have argued that the separated rate matrix $\Gamma^{S}$ is physical, at least in the case that the original noise is always PSD. We have also found that the strength of this separated noise is the same as the strength of the original noise, whenever the original noise is constant. Therefore, in such a situation, the transformation which maps $\Gamma^{D}$ to $\Gamma^{S}$ is necessarily a positive, trace-preserving map. Despite the strength of the analytic arguments presented above, we also present additional numerical evidence of this claim. We will focus here on proving this claim in the fully time-independent case, in which we can use the transformation
\begin{equation}
\Gamma^{S} ~=~ \mathcal{K} \left [ \Gamma^{D} \right ] ~=~ \sum_{m=0}^{\infty} \frac{t_{\text{op}}^{m}}{(m+1)!} \xi^{m} \left [ \Gamma^{D} \right ].
\end{equation}

In order to verify that our transformation is a positive, trace-preserving map, we will focus on demonstrating that the associated Choi matrix of the map is positive, where the Choi matrix is defined according to
\begin{equation}
C_{\mathcal{K}} = \sum_{i,j} e^{ij} \otimes \mathcal{K} \left [ e^{ij} \right],
\end{equation}
where $e^{ij}$ is the matrix with $(i, j)$ entry equal to one, with zeros elsewhere. We emphasize that since the transformation $\mathcal{K}$ is a transformation on the space of rate matrices, the matrices $ \left \{ e^{ij} \right \}$ have dimensions $\left ( \mathcal{D}^{2}-1 \right ) \times \left ( \mathcal{D}^{2}-1 \right )$. Proving that the Choi matrix is positive would in fact actually prove a somewhat stronger claim, which is that the map is not only positive, but rather completely positive (it is still positive after being tensored with the identity map of arbitrary dimensionality). Interestingly, this does indeed seem to be true for all of the systems that we have studied numerically. For this reason, we suspect that the transformation, at least in the time-independent case, is indeed always completely positive, rather than simply positive. However, since this fact is not needed for ensuring the physical nature of $\Gamma^{S}$, we do not make an effort to prove this claim in the general case. We simply note that the positivity of the Choi matrix is more than sufficient to prove the positivity of the map, and so it serves as a sufficient numerical test of our results.

While proving the positivity of the Choi matrix in the general case is a formidable task, it is a relatively straight-forward exercise to compute this matrix numerically, for specific choices of the Hamiltonian. In order to compute the Choi matrix, we must numerically evaluate the action of $\mathcal{K}$ on the various matrices $e^{ij}$, using the methods outlined in Appendix~\ref{sec:appNumerics}. Each choice of Hamiltonian will result in a different transformation $\xi$, and thus a different Choi matrix. Having evaluated the Choi matrix for all single-qubit and pair-interaction Hamiltonian terms, we have found it to possess a positive spectrum in all such cases. Since we assume that the operations $ \left \{ \phi^{(i)} \right \}$ composing the coherent part of a quantum gate are likely to consist primarily of one and two-body interaction terms, this provides a numerical sanity check for the physical nature of the separated noise $\mathcal{L}_{S}$ in most relevant cases of interest, at least in the time-independent case.

Interestingly, for the case of a single-qubit term, it is in fact possible to derive an analytic expression for the Choi matrix and its spectrum. Let us consider an operation with
\begin{equation}
\mathcal{L} \left [ \rho \right ] = -i \left [ H, \rho \right ] + \sum_{\alpha, \beta} \Gamma^{D}_{\alpha \beta} \left [ \sigma^{\alpha} \rho \sigma^{\beta} - \frac{1}{2} \left \{ \sigma^{\beta} \sigma^{\alpha}, \rho \right \} \right ],
\end{equation}
where the $\left\{\sigma^{\alpha}\right\}$ are the usual Pauli operators acting on a single qubit. Here we choose to take
\begin{equation}
H = - J \sigma^{x},
\end{equation}
for some positive $J > 0$. Any other choice of Hamiltonian is equivalent through rotational symmetry. The rate matrix in this case is a $3\times3$ positive semi-definite matrix, 
\begin{equation}
\Gamma^{D} \to 
\begin{bmatrix}
   \Gamma^{D}_{XX} & \Gamma^{D}_{XY} & \Gamma^{D}_{XZ} \\
   \Gamma^{D}_{YX} & \Gamma^{D}_{YY} & \Gamma^{D}_{YZ} \\
   \Gamma^{D}_{ZX} & \Gamma^{D}_{ZY} & \Gamma^{D}_{ZZ}
   \end{bmatrix}
\end{equation}
while the Choi matrix is a $9\times9$ matrix, constructed according to the definition above. While the full Choi matrix is too cumbersome to reproduce in its entirely here, its entries consist of relatively simple trigonometric functions of the operation angle $\theta$, where we again define
\begin{equation}
\theta = 2 J t_{\text{op}}
\end{equation}

\begin{figure}
   \centering
   \includegraphics[width=0.45\textwidth]{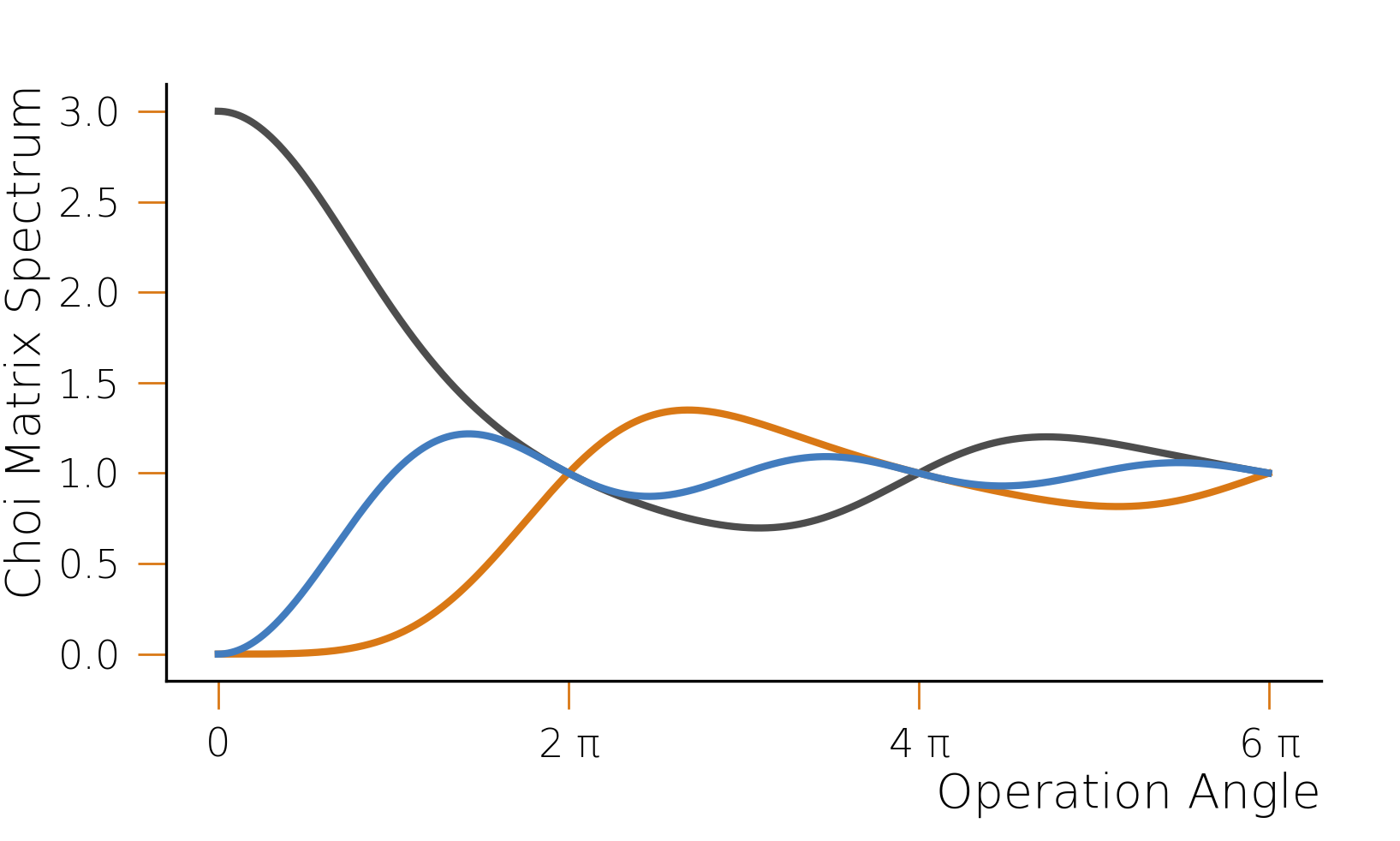}
   \caption{The spectrum of the Choi matrix for the single-qubit case, as a function of the operation angle $\theta$.}
   \label{fig:choi}
\end{figure}

Much more interesting than the structure of the Choi matrix is its spectrum, which can also be computed in closed form. The Choi matrix in this case possesses three non-zero eigenvalues, which are given according to
\begin{equation}
\begin{split}
& \lambda_{0} = 1-\frac{\sin (\theta)}{\theta} ~;~ \\
& \lambda_{\pm} = 1 +  \frac{ \sin (\theta) \pm \sqrt{2\cos (\theta)+34} \sin \left(\theta / 2\right)}{2 \theta}
\end{split}
\end{equation}
A plot of this spectrum is shown in Figure~\ref{fig:choi}. The positive nature of the spectrum demonstrates explicitly that the Choi matrix is positive, and hence the map $\mathcal{K}$ is a (completely) positive map in the single-qubit case.

%% file: Sections/appNumerics.tex
\section{Evaluation of the Separated Noise in the Time-Independent Case}
\label{sec:appNumerics}

Here we briefly outline how we find the separated noise in the time-independent case. In order to compute the separated noise, we must evaluate the action of $\mathcal{K}$ on the rate matrix $\Gamma^{D}$ of the original noise,
\begin{equation}
\Gamma^{S} = \mathcal{K} \left [ \Gamma^{D} \right] = \sum_{m=0}^{\infty} \frac{ t_{\text{op}}^{m}}{(m+1)!} \xi^{m} \left [ \Gamma^{D} \right ],
\end{equation}
To do so, we first note that the series expansion defining $\mathcal{K}$ can in fact be summed exactly, resulting in the expression
\begin{equation}
f \left ( x \right ) = \sum_{m=0}^{\infty} \frac{x^{m}}{(m+1)!} = \frac{e^{x}-1}{x}.
\end{equation}
We emphasize that despite the appearance of this expression with a power of $x$ in the denominator, it is indeed analytic. With this, we can write
\begin{equation}
\mathcal{K} = f \left ( t_{\text{op}} \xi \right ).
\end{equation}

Secondly, we note that $\xi$, acting as a transformation on the space of rate matrices, is in fact anti-Hermitian with respect to the Frobenius inner product, 
\begin{equation}
\text{Tr}\left [ \left ( \Gamma^{A} \right )^{\dagger} \xi \left [ \Gamma^{B} \right ] \right ] = -\text{Tr}\left [ \left ( \xi \left [ \Gamma^{A} \right ] \right )^{\dagger}  \Gamma^{B}  \right ]
\end{equation}
Hence, if we evaluate the matrix $\mathcal{M}_{\xi}$ of the transformation $\xi$ in some basis which is orthonormal with respect to this norm, then this matrix will itself be anti-Hermitian. It is therefore always possible to diagonalize $\mathcal{M}_{\xi}$,
\begin{equation}
\mathcal{M}_{\xi} = \mathcal{U} \Delta_{\xi} \mathcal{U}^{\dagger}
\end{equation}
for some diagonal $\Delta_{\xi}$ and unitary $\mathcal{U}$. We note that for the case of rate matrices, we choose to define the Frobenius inner product \textbf{without} any factors involving the Hilbert space dimension $\mathcal{D}$, which is in contrast with our definition for the space of traceless operators.

With these two facts, we see that in order to evaluate the full transformation $\mathcal{K}$ on the space of rate matrices, one can compute the matrix $\mathcal{M}_{\xi}$ in some orthonormal basis of rate matrices, diagonalize this matrix to find $\Delta_{\xi}$, compute the function $f$ on the resulting diagonal matrix, and rotate back to the original basis in order to find the matrix $\mathcal{M}_{\mathcal{K}}$ corresponding to the full transformation $\mathcal{K}$. That is to say, we can find $\mathcal{K}$ according to
\begin{equation}
\mathcal{M}_{\mathcal{K}} = \mathcal{U} f \left ( t_{\text{op}} \Delta_{\xi} \right )  \mathcal{U}^{\dagger} = \mathcal{U} f \left ( t_{\text{op}} \mathcal{U}^{\dagger} \mathcal{M}_{\xi}  \mathcal{U} \right )  \mathcal{U}^{\dagger}.
\end{equation}
The transformation $\mathcal{K}$ can then be applied to an arbitrary rate matrix by writing this rate matrix as a vector in our chosen basis and applying the matrix $\mathcal{M}_{\mathcal{K}}$ to it. For simplicity, we choose to work with the (orthonormal) basis $\left \{ e^{ij} \right \}$, where $e^{ij}$ is the matrix with $(i, j)$ entry equal to one, with zeros elsewhere. However, other choices are of course possible.

We note that instead of rotating the object $f \left ( t_{\text{op}} \Delta_{\xi} \right )$ back to the original basis to find $\mathcal{M}_{\mathcal{K}}$, we could instead express the rate matrix $\Gamma^{D}$ as a vector in the eigenbasis of $\mathcal{M}_{\xi}$, and then apply the object $f \left ( t_{\text{op}} \Delta_{\xi} \right )$ directly to this vector. This is the approach we take when studying the steady state of the separated noise in Appendix~\ref{sec:appSteady}.

Interestingly, for the case of a single-qubit Hamiltonian, it is in fact possible to derive an analytic expression for the matrix $\mathcal{M}_{\xi}$ and its spectrum, and thus the separated noise. In this single qubit case, we perform the above steps using Mathematica to arrive at an analytic expression for the matrix $\mathcal{M}_{\mathcal{K}}$ of the full transformation as a function of the operation angle of the Hamiltonian. We can then apply this transformation to various choices of hardware noise, in order to arrive at the expressions found in Section~\ref{sec:sepNoiseExamp} of the main text. 

For the case of a single qubit, directly computing the spectrum of $\xi$ is easy enough that it could, in principle, be done by hand (we see in Appendix~\ref{sec:appSteady} how this is tantamount to studying the algebra of the Gell-Mann matrices). However, as we increase the number of qubits in our system of interest, the dimensionality of the transformation $\xi$ grows dramatically. In particular, since the transformation $\xi$ acts on the space of $\left ( \mathcal{D}^{2}-1 \right ) \times \left ( \mathcal{D}^{2}-1 \right )$ rate matrices, the vector space on which $\xi$ acts has a dimensionality which grows like the fourth power of the Hilbert space dimension $\mathcal{D}$ of the Hamiltonian. Thus, diagonalizing the transformation $\xi$ is computationally feasible for systems composed of at most a handful of qubits. 

Fortunately, however, finding the spectrum of the transformation $\xi$ does not require such a direct approach. Since the transformation $\xi$ is the adjoint action under the matrix $\Omega$, which is itself (the matrix representation of) the adjoint action of the Hamiltonian, it is possible to build the spectrum of $\xi$ directly from the spectrum of the Hamiltonian. While diagonalizing the Hamiltonian of a many-body system may also constitute a formidable computational problem in the general case, we can still hope to access significantly larger system sizes using this indirect approach.

To see how we can accomplish this in practice, we demonstrate such a process for the case of a single qubit (though we emphasize that such a process is most useful when considering a many-qubit system). The energy eigenstates of the Hamiltonian are given according to
\begin{equation}
H | \pm \rangle = E_{\pm}| \pm \rangle .
\end{equation}
From these energy eigenstates, we construct the operators
\begin{equation}
B_{\pm} = \sqrt{2}| \pm \rangle \langle \mp | ~;~ B_{0} = |+\rangle \langle + | - |-\rangle \langle - |
\end{equation}
It is straightforward to verify that these operators form an orthonormal basis for the space of all traceless operators acting on the original Hilbert space, and so we can express the Lindblad dissipator using this basis,
\begin{equation}
\mathcal{L}_{D}\left [ \rho  \right ] = \sum_{n, m} \Gamma^{D;B}_{nm} \left ( B_{n} \rho B_{m}^{\dagger} - \frac{1}{2} \left \{ B_{m}^{\dagger} B_{n}, \rho \right \}  \right ).
\end{equation}
We recall that if this basis $ \left \{ B_{n} \right \} $ constitutes a new basis with respect to some old basis $ \left \{ A_{n} \right \} $ in which we had originally written the dissipator, such that
\begin{equation}
B_{n} = \sum_{p} V_{pn} A_{p} ~;~ V_{pn} = \frac{1}{2}\text{Tr} \left [ A_{p}^{\dagger} B_{n} \right ] ,
\end{equation}
then the rate matrices with respect to these two bases are related according to
\begin{equation}
\Gamma^{D;A} = V \Gamma^{D;B} V^{\dagger}.
\end{equation}
In particular, the matrix elements of $V$ can be found in terms of matrix elements of the operators $ \left \{ A_{n} \right \} $ in the eigenbasis of the Hamiltonian. For example,
\begin{equation}
V_{p+} = \frac{1}{\sqrt{2}}\text{Tr} \left [ A_{p}^{\dagger} |+\rangle \langle -| \right ] =  \frac{1}{\sqrt{2}}\langle - | A_{p}^{\dagger} | + \rangle.
\end{equation}

The utility of this new basis $ \left \{ B_{n} \right \} $ can be seen by noting that these operators are eigenmodes of the adjoint action of the Hamiltonian, such that
\begin{equation}
\text{ad}_{H} \left [ B_{\pm} \right ] = \pm \left ( E_{+} - E_{-} \right ) B_{\pm} ~;~ \text{ad}_{H} \left [ B_{0} \right ] = 0.
\end{equation}
The matrix $\Omega$ is therefore diagonal in this basis, with matrix elements given by
\begin{equation}
\Omega_{\pm \pm} = \pm \left ( E_{+} - E_{-} \right ) ~;~ \Omega_{0 0} = 0.
\end{equation}
With this information, it is easy to verify that the eigenmatrices of the transformation $\xi$ are given according to
\begin{equation}
\xi \left [ e^{ij} \right ] = -i \left ( \Omega_{ii} - \Omega_{jj} \right ) e^{ij},
\end{equation}
where $e^{ij}$ is the matrix with $(i, j)$ entry equal to one, and zeros elsewhere. We thus have an explicit construction for the spectrum of $\xi$ in terms of the spectrum of the Hamiltonian.

In the case of a general quantum system with Hilbert space dimension $\mathcal{D}$, the construction of the $B$ operators proceeds in a fashion similar to the construction of the generalized Gell-Mann matrices. If the energy eigenstates of the Hamiltonian are given according to
\begin{equation}
H | i \rangle = E_{i}| i\rangle,
\end{equation}
then the generalization of the $B_{\pm}$ operators is given by the operators
\begin{equation}
B_{ij} = \sqrt{\mathcal{D}}| i \rangle \langle j |,
\end{equation}
while the generalization of the $B_{0}$ operator is given by the collection of operators
\begin{equation}
B_{0k} = \frac{\sqrt{\mathcal{D}}}{\sqrt{k\left(k+1\right)}} \left ( \sum_{m=1}^{k} | m \rangle \langle m | - k | k+1 \rangle \langle k+1 | \right ),
\end{equation}
with
\begin{equation}
1 \leq k \leq \mathcal{D} - 1.
\end{equation}
With these operators, the construction of the spectrum of $\xi$ proceeds in a fashion analogous to the case of a two-level system.

%% file: Sections/appSteady.tex
\section{Steady State of the Separated Noise}
\label{sec:appSteady}

Here we derive the form of the steady state separated noise in the time-independent case, and also discuss the angular scale over which the separated noise reaches this steady state.

Since we consider the time-independent case, we can use the transformation 
\begin{equation}
\Gamma^{S} ~=~ \frac{e^{t_{\text{op}} \xi}-\mathcal{I}}{t_{\text{op}} \xi} \left [ \Gamma^{D} \right ] 
\end{equation}
to compute the rate matrix of the separated noise from the rate matrix of the original noise. As noted in Appendix~\ref{sec:appNumerics}, the transformation $\xi$ is anti-Hermitian, and so its eigenvectors form a complete basis for the space of rate matrices. We can therefore decompose the original noise as
\begin{equation}
\Gamma^{D} ~=~ \sum_{a} \gamma_{a} \Gamma_{a},
\end{equation}
with
\begin{equation}
\xi \left [ \Gamma_{a} \right ] = 2 J i \eta_{a} \Gamma_{a}
\end{equation}
where $\eta_{a}$ is strictly real and $J$ is the characteristic energy scale of the Hamiltonian. Applying the separated noise transformation to this expression, we find
\begin{equation}
\Gamma^{S} = \sum_{a} \gamma_{a} \left ( \theta \right ) \Gamma_{a} = \sum_{a} \frac{\gamma_{a}}{i \eta_{a} \theta} \left ( e^{i \eta_{a} \theta } - 1 \right ) \Gamma_{a}
\end{equation}
where we have again defined
\begin{equation}
\theta = 2 J t_{\text{op}}
\end{equation}

To understand the nature of the steady state separated noise, we analyze the evolution of each component of the separated noise in the eigenbasis of $\xi$, according to its associated eigenvalue $\eta_{a}$. When $\eta_{a}$ is zero, we have a component in the null space of the operator $\xi$. Such a component is in fact static, since
\begin{equation}
\frac{1}{i \eta_{a} \theta} \left ( e^{i \eta_{a} \theta} - 1 \right ) \to 1
\end{equation}
for any finite $\theta$, in the limit that $\eta_{a}$ is zero. In contrast to this, when $\eta_{a}$ is not zero, the term $\gamma_{a} \left ( \theta \right )$ will exhibit decaying oscillations as a function of the operation angle, eventually approaching zero. At large operation angles, only the projection of the underlying hardware noise into the null space of the transformation $\xi$ will remain, thereby identifying this object as the steady state of the separated noise. We will refer to the portion of the separated noise which does evolve (and which eventually decays to zero) as the residual separated noise. 

The angular scale over which a given component of the residual separated noise decays to some fraction of its initial value is set entirely by the (non-zero) eigenvalue $\eta_{a}$,
\begin{equation}
\frac{ \gamma_{a} \left ( \theta \right )}{ \gamma_{a} \left ( 0 \right )} = \frac{1}{i \eta_{a} \theta } \left ( e^{i \eta_{a} \theta } - 1 \right ).
\end{equation}
Since the eigenvalues $\eta_{a}$ are determined solely by the dimensionless coherent dynamics, the angular rate at which a given component of the residual separated noise decays is independent of both the energy scale of the Hamiltonian and the noise strength of the original noise. However, the particular form of the separated noise does affect which eigenmodes it couples to, and these eigenmodes are further determined by the (dimensionless) coherent dynamics of the system. Since the angular scale for the approach of the separated noise to its steady state value is dictated by the slowest eigenmode which the original noise couples to, we see that this time scale is independent of both the overall energy scale of the Hamiltonian and the overall noise strength of the original noise, but it is affected by the interplay between the form of the original noise and the coherent dynamics.

We note that an individual component of the residual separated noise will vanish at any finite operation angle which satisfies
\begin{equation}
e^{i \eta_{a} \theta } - 1 = 0.
\end{equation}
If this relationship is satisfied at a given angle for every component simultaneously, then the residual separated noise will reach its steady state value at this finite operation angle.

To study this approach to the steady state in more detail, we again consider the example of dephasing noise from Section~\ref{sec:sepNoiseExamp}. The transformation $\xi$ in this case is given according to
\begin{equation}
\xi = -i \text{ad}_{\Omega} = 2 J i \left [ \lambda_{7}, \cdot \right ],
\end{equation}
where $\lambda_{7}$ is the seventh Gell-Mann matrix. In the main text, we have discussed the null space of this transformation, along with the projection of the original noise into this null space (thus giving the steady state of the separated noise). Here we focus on the residual separated noise. It is straightforward to verify that
\begin{equation}
\xi \left [ \Gamma_{a} \right ] = 2 J i \eta_{a} \Gamma_{a} ~ \Leftrightarrow ~ \xi \left [ \Gamma_{a}^{\dag} \right ] = -2 J i \eta_{a} \Gamma_{a}^{\dag},
\end{equation}
indicating that the eigenmodes of $\xi$ come in conjugate pairs (this is in fact a generic feature of the transformation $\xi$, not specific to this example). Likewise, the components of the original noise along these eigenmodes obey the relationship
\begin{equation}
\gamma_{\overline{a}} = \langle \langle  \Gamma_{a}^{\dag} || \Gamma^{D} \rangle \rangle = \langle \langle  \Gamma_{a} || \Gamma^{D} \rangle \rangle^{*} = \gamma_{a}^{*}
\end{equation}
In the case of our example, the only two eigenmodes with non-zero eigenvalue which couple to the original noise are given (with proper normalization) according to
\begin{equation}
\Gamma_{\pm} = -\frac{1}{4} \lambda_{3} \pm \frac{1}{2}i\lambda_{6} + \frac{\sqrt{3}}{4} \lambda_{8},
\end{equation}
with corresponding eigenvalues
\begin{equation}
\eta_{\pm} = \pm 2.
\end{equation}
The coupling to these modes is given by
\begin{equation}
\gamma_{\pm} = -\frac{1}{4} \gamma_{\text{deph}}.
\end{equation}
Thus, the components of the residual separated noise evolve according to
\begin{equation}
\gamma_{\pm} \left ( \theta \right ) = \pm \frac{\gamma_{\text{deph}}}{8 \theta} i \left ( e^{ \pm 2 i \theta } - 1 \right ).
\end{equation}
We note that both of these components vanish at integer multiples of $\pi$, and so the residual separated noise will reach its steady state value at these finite angles. 

For a generic many-body system, the angular scale over which the separated noise will approach its steady state value will depend on the spectrum of the (dimensionless) transformation $\xi$, as well as the relationship between its eigenmodes and the original noise. As mentioned in Appendix~\ref{sec:appNumerics}, the eigenvalues of the transformation $\xi$ will be given by differences of differences of energy eigenvalues of the Hamiltonian. Since we expect a generic many-body system to possess an average energy splitting which becomes exponentially small in the size of the system, there will exist eigenvalues $\eta_{a}$ which become arbitrarily small in this limit. It may then naively appear that the angular relaxation scale of the separated noise will diverge as we increase the size of the system. However, this relaxation scale will also depend on whether the original noise couples to these slow eigenmodes. For a generic, non-integrable Hamiltonian, the structure of these eigenmodes will be non-trivial, and making definitive statements about this relaxation scale represents a significant avenue of research in its own rite, which is beyond the scope of this work.

%% file: Sections/appGateExamp.tex
\section{A Concrete Example of a Noise Term Following a Complete Gate Operation}
\label{sec:appGateExamp}

Here we provide a derivation of the gate noise example given in the main text. In particular, we consider a toy model of an ideal Hadamard gate, whose unitary time-evolution matrix in the noise-free case would correspond to
\begin{equation}
U \to \exp(-i\frac{\pi}{4}Y)\exp(+i\frac{\pi}{2}Z).
\end{equation}
Our toy model thus corresponds to a sequence of two elementary rotations. We note that due to the simplicity of our chosen toy model, there are some hardware contexts in which it may be more natural to think of the above manipulations as instead being a sequence of two proper gates, rather than one gate composed of two operations. While this may indeed be the case for our example, we emphasize that more complicated gates, especially those involving two qubits, will generally involve elementary operations which one would not interpret as corresponding to the native gate set of the hardware, especially when these operations involve time-dependent Hamiltonians. In any case, we are only concerned here with providing a basic illustrative example.

In addition to the coherent evolution above, we assume that there is hardware noise occurring during the implementation of this gate, which in this case we take to be damping noise,
\begin{equation}
\mathcal{L}_{\text{damp}} \left [ \rho \right ] =  \gamma_{\text{damp}}  \left [ \sigma^{+} \rho \sigma^{-} - \frac{1}{2} \left \{\sigma^{-}\sigma^{+}, \rho \right \}  \right ].
\end{equation}
We wish to understand how this noise alters the behavior of the otherwise noise-free Hadamard gate. To do so, we must compute the effective noise model according to our previously outlined approach. That is to say, we must first determine the separated noise for each elementary operation, then commute all noise terms to the left, and finally sum up the commuted noise.

To compute the separated noise for the first elementary operation, we must consider the Lindblad evolution
\begin{equation}
\begin{split}
& \mathcal{L}_{Z} \left [ \rho \right ] = \\
& -i J_{Z} \left [ \sigma^{z}, \rho \right ] +  \gamma_{\text{damp}}  \left [ \sigma^{+} \rho \sigma^{-} - \frac{1}{2} \left \{\sigma^{-}\sigma^{+}, \rho \right \}  \right ]
 \end{split}
\end{equation}
for some appropriate coupling $J_{Z}$. In this case, we find that the matrix $\Omega_{Z}$ corresponding to the coherent evolution commutes with the rate matrix for damping noise
\begin{equation}
\left [ \Omega_{Z}, \Gamma^{D} \right ] =  0,
\end{equation}
and so in this case, there is no change to the noise upon separation. We thus find simply
\begin{equation}
\Gamma^{S}_{Z} ~=~
\frac{1}{2}\gamma_{\text{damp}}
\begin{bmatrix}
  \frac{1}{2} & -\frac{i}{2} & 0 \\
  \frac{i}{2} & \frac{1}{2} & 0 \\
  0 & 0 & 0 \\
\end{bmatrix}.
\end{equation}
For the second operation, we similarly write
\begin{equation}
\begin{split}
& \mathcal{L}_{Y} \left [ \rho \right ] = \\
& -i J_{Y} \left [ \sigma^{y}, \rho \right ] +  \gamma_{\text{damp}}  \left [ \sigma^{+} \rho \sigma^{-} - \frac{1}{2} \left \{\sigma^{-}\sigma^{+}, \rho \right \}  \right ]
\end{split}
\end{equation}
for some appropriate coupling $J_{Y}$. We note that this coupling satisfies
\begin{equation}
2 J_{Y} t^{Y}_{\text{op}} \equiv \theta^{Y}  = \pi / 2.
\end{equation}
Computing the transformation $\mathcal{K}$ on the underlying rate matrix for this second coherent evolution, we find the separated noise to be
\begin{equation}
\Gamma^{S}_{Y} ~=~
\frac{1}{2}\gamma_{\text{damp}}
\begin{bmatrix}
  \frac{1}{4} & -\frac{i}{\pi} & -\frac{1}{2\pi} \\
 \frac{i}{\pi} & \frac{1}{2} & -\frac{i}{\pi} \\
 -\frac{1}{2\pi} & \frac{i}{\pi} & \frac{1}{4} \\
\end{bmatrix}.
\end{equation}
We note that this separated noise looks significantly different from the original damping noise.

Having found the separated noise for each operation, we must now commute all of the noise terms to the left of the coherent terms. For the first operation, we must move the separated noise past the coherent piece of the second operation (the rotation around the y axis). Using our previously stated results for the transformation of a noise term after being commuted to the left of a coherent term, we find that the separated noise of the first operation, after being commuted past the coherent gate, is
\begin{equation}
\Gamma^{L}_{Z} ~=~
\frac{1}{2}\gamma_{\text{damp}}
\begin{bmatrix}
  0 & 0 & 0 \\
  0 & \frac{1}{2} & -\frac{i}{2} \\
  0 & \frac{i}{2} & \frac{1}{2} \\
\end{bmatrix}.
\end{equation}
Notice that the noise has essentially been ``rotated'' around the y axis. 

For the second operation, the separated noise term is already to the left of all coherent operations, and thus no transformation is necessary. We therefore have simply
\begin{equation}
\Gamma^{L}_{Y} ~=~
\frac{1}{2}\gamma_{\text{damp}}
\begin{bmatrix}
  \frac{1}{4} & -\frac{i}{\pi} & -\frac{1}{2\pi} \\
 \frac{i}{\pi} & \frac{1}{2} & -\frac{i}{\pi} \\
 -\frac{1}{2\pi} & \frac{i}{\pi} & \frac{1}{4} \\
\end{bmatrix}.
\end{equation}

Now that all of the noise terms have been commuted to the left, we can add them together to zero order in the BCH formula,
\begin{equation}
t_{G}\mathcal{L}_{N} = t^{Z}_{\text{op}} \mathcal{L}^{Z}_{L} + t^{Y}_{\text{op}} \mathcal{L}^{Y}_{L}
\end{equation}
or, in terms of the underlying rate matrices,
\begin{equation}
\Gamma^N = \frac{t^{Z}_{\text{op}}}{t_{G}} \Gamma^{L}_{Z} + \frac{t^{Y}_{\text{op}}}{t_{G}} \Gamma^{L}_{Y}
\end{equation}
where we have defined 
\begin{equation}
t_{G} = t^{Z}_{\text{op}} + t^{Y}_{\text{op}}.
\end{equation}
Since the first operation implements an angle which is twice as large as the second operation, we may imagine a hypothetical situation in which
\begin{equation}
t^{Z}_{\text{op}} = 2 t^{Y}_{\text{op}}.
\end{equation}
In this case, we can write
\begin{equation}
\Gamma^N = \frac{2}{3} \Gamma^{L}_{Z} + \frac{1}{3} \Gamma^{L}_{Y}.
\end{equation}
Explicitly, this is
\begin{equation}
\Gamma^{N} ~=~
\frac{1}{2}\gamma_{\text{damp}}
\begin{bmatrix}
\frac{1}{12} & -\frac{i}{3 \pi } & -\frac{1}{6 \pi } \\
 \frac{i}{3 \pi } & \frac{1}{2} & -\frac{i (1+\pi )}{3 \pi } \\
 -\frac{1}{6 \pi } & \frac{i (1+\pi )}{3 \pi } & \frac{5}{12} \\
\end{bmatrix}
\end{equation}
or, evaluated numerically,
\begin{equation}
\Gamma^{N} \approx
\frac{1}{2}\gamma_{\text{damp}}
\begin{bmatrix}
0.083 & -0.106 i & -0.053 \\
0.106i & 0.5 & -0.439 i \\
 -0.053 & 0.439 i & 0.417 \\
\end{bmatrix}.
\end{equation}

Despite the relatively simple nature of both the underlying hardware noise and the coherent gate, the effective noise term can take a relatively complex form. We again emphasize that we have arrived at this result without making any assumptions about the magnitude of $\gamma_{\text{damp}}$, or its relationship to the operation times $t^{Z}_{\text{op}}$ and $t^{Y}_{\text{op}}$.

We note that in the above derivation, we have assumed that the elementary operations are performed immediately one after another, without any idling time in between. However, it is of course entirely possible that in an actual hardware implementation, significant idling time may be required between each elementary operation. In such a case, our ultimate expression for the gate noise would satisfy
\begin{equation}
t_{G}\mathcal{L}_{N} = t^{Z}_{\text{op}} \mathcal{L}^{Z}_{N} + t^{Y}_{\text{op}} \mathcal{L}^{Y}_{N} + t_{\text{idle}} \mathcal{L}_{\text{damp}}.
\end{equation}
If we measure the idling time in units of $t^{Y}_{\text{op}}$,
\begin{equation}
t_{\text{idle}} = m t^{Y}_{\text{op}},
\end{equation}
and thus associate
\begin{equation}
t_{G} = \left ( m+3 \right )t^{Y}_{\text{op}},
\end{equation}
we now find
\begin{equation}
\Gamma^{N} = \frac{2}{m+3} \Gamma^{L}_{Z} + \frac{1}{m+3} \Gamma^{L}_{Y} + \frac{m}{m+3} \Gamma^{\text{damp}}.
\end{equation}
In cases in which the idling time greatly exceeds the time required for a single rotation, we have
\begin{equation}
m \gg 1,
\end{equation}
and we thus find
\begin{equation}
\Gamma^{N} \approx \Gamma^{\text{damp}}.
\end{equation}

We thus see that while it is indeed possible for the noise term following a gate to differ significantly from the underlying hardware noise, this may not always be the case. In particular, when the time required for elementary operations is significantly less than the idling time between them, we should expect the gate noise to still be dominated by the underlying hardware noise.

However, we emphasize that the requirement stated above is that the idling time occurring in between elementary operations is large compared with the duration of these individual operations themselves. This requirement does \textit{not} place any restrictions on how these operation and idling times compare with respect to the time scales of the underlying hardware noise. The individual rate matrices $\Gamma^{L}_{Z}$ and $\Gamma^{L}_{Y}$ will differ significantly from the underlying hardware noise whenever their associated operation angles are large, regardless of whether their associated operation times are small with respect to the time scales of the underlying hardware noise. It is also irrelevant how long the total gate time is, since many individual operations executed in rapid succession with minimal idling time between them can still result in noise which differs significantly from the underlying hardware noise, even if the total time required to perform all of these operations is ``small'' in whatever appropriate sense one wishes to define.